\documentclass[preprint,twocolumn]{aastex631}

\usepackage{subfigure}
\usepackage{color}
\usepackage{xcolor}
\usepackage{amsmath}
\usepackage{mathrsfs}
\usepackage{multirow}
\usepackage{soul}
\usepackage{amssymb}
\usepackage{hyperref}
\usepackage{threeparttable}

\definecolor{ultramarine}{rgb}{0.8, 0.1, 0.4}

\def\be{\begin{equation}}
\def\ee{\end{equation}}
\def\bd{\begin{displaymath}}
\def\ed{\end{displaymath}}
\def\ba{\begin{aligned}}
\def\ea{\end{aligned}}

\def\nms{\mathsurround=0pt}

\def\oversim#1#2{\lower 4pt\vbox{\baselineskip 0pt \lineskip 1pt
    \ialign{$\nms#1\hfil##\hfil$\crcr#2\crcr\sim\crcr}}}

\def\bh{M_{\bullet}}

\def\msun{M_{\odot}}

\def\pl{\partial}

\def\GNC{\texttt{GNC}}
\def\GNCp{\texttt{GNC(v1)}}
\def\GNCc{\texttt{GNC(v2)}}

\begin{document}

\title{Self-consistent Solutions of  Evolving Nuclear Star Clusters with Two-Dimensional Monte-Carlo Dynamical Simulations}
\author{Fupeng Zhang}
\correspondingauthor{FUPENG ZHANG}
\affiliation{School of Physics and Materials Science, Guangzhou
University, Guangzhou 510006, China}[0]
\email{zhangfupeng@gzhu.edu.cn}
\affiliation{Key Laboratory for Astronomical Observation and Technology of Guangzhou, Guangzhou 510006, China}
\affiliation{Astronomy Science and Technology Research Laboratory of Department of Education of Guangdong Province, Guangzhou 510006, China}
\author{Pau Amaro Seoane}
\affiliation{Institute of Multidisciplinary Mathematics, Universitat Politècnica de València, Spain}
\affiliation{
Max-Planck-Institute for Extraterrestrial Physics, Garching, Germany}
\affiliation{
The Higgs Centre for Theoretical Physics, University of Edinburgh, UK
}

\begin{abstract}
We recently developed a Monte-Carlo method (\GNC) that can simulate the dynamical evolution of a nuclear stellar cluster (NSC) with a massive black hole (MBH), where the two-body relaxations can be solved by the Fokker-Planck equations in energy and angular momentum space.  Here we make a major update of \GNC~ by integrating stellar potential and adiabatic invariant theory, so that we can study the self-consistent dynamics of NSCs with increasing mass of the MBH.  We perform tests of the self-adaptation of cluster density due to MBH mass growth and Plummer core collapse, both finding consistent results with previous studies, the latter having a core collapse time of $\sim 17t_{\rm rh}$ by \GNC, where $t_{\rm rh}$ is the time of half-mass relaxation.  We use \GNC~ to study the cosmological evolution of the properties of NSC and the mass of MBH assuming that the mass growth of the MBH is due to loss-cone accretion of stars (e.g., tidal disruption of stars) and stellar black holes,  and compare the simulation results with the observations of NSCs in Milky-Way or near-by galaxies.
Such scenario is possible to produce MBHs with mass $10^5\sim 10^7\,\msun$ for NSCs with stellar mass of $10^6\sim 10^9\,\msun$. In Milky-Way's NSC, to grow MBH up to $4\times 10^6\,\msun$, 
its size needs to be $\sim 1.7$ times more compact in early universe than the current value. MBHs with current masses $>6\times 10^{7}\,\msun$ seem difficult to explain by loss-cone accretion alone, and thus may require other additional accretion channels, such as gas accretion. 
\end{abstract}


\section{Introduction}

The very dense star clusters that contain a central massive black hole (MBH) at the centre of almost all galaxies, i.e. the generally called ``nuclear star clusters (NSCs)''\footnote{though we note that observationally NSCs usually means the stellar component that can be distinguished from the bulge components of the galaxy (to identify them).  Here we use it in a more general way to infer all stellar components around an MBH, up to the length scales of interest.}, are the epicentre of the cosmological formation and evolution of MBHs~\citep[e.g., See also reviews in~\citet{2020A&ARv..28....4N}]{1987ApJ...321..199Q, 2002A&A...394..345F, 2006MNRAS.368..141F, 2006ApJ...649...91F, 2002ApJ...576..899P, 2010MNRAS.407.1529H, 2017MNRAS.467.4180S, 2020ARA&A..58...27I}, and the dynamical formation of a number of gravitational wave sources~\citep[e.g.,][]{1997MNRAS.284..318S, 2003ApJ...583L..21F,2005ApJ...629..362H,2013MNRAS.429.3155A,2014MNRAS.437.1259B, 2018LRR....21....4A,2019PhRvD..99l3025A,2011CQGra..28i4017A, AmaroSeoane17, 2022hgwa.bookE..17A,1987ApJ...321..199Q,Oleary09, 2015MNRAS.448..754H, Antonini12,Antonini16, Wen03, Zhang19,Zhang21, 2017ApJ...835..165B,2018CmPhy...1...53C,2023PhRvD.107d3009X}.  It also provides a natural laboratory for testing gravitational theories in strong fields, such as those at the centre of our own galaxy\citep[e.g.,][]{2010PhRvD..81f2002M, 2015ApJ...809..127Z,2016CQGra..33k3001J,2016ApJ...827..114Y,2019MNRAS.489.4606G,2018A&A...615L..15G}.  There is therefore a strong motivation to develop numerical methods that can accurately follow the evolution of each stellar component and of the cluster as a whole, in order to provide accurate predictions of these important astrophysical questions. 

However, the secular dynamical evolution of NSCs with a central MBH is a challenging problem for theoretical astrophysics. The dynamics is mainly driven by two-body relaxation, which is highly nonlinear, multidimensional and spans a wide range of scales in both the length and period of stellar orbits~\citep[e.g.,][]{1972PhRvD...5.1021H,SM78,1978ApJ...226.1087C}.  The situation is further complicated by the intense interactions between different types of stellar objects and the central MBH, such as tidal disruption of stars~\citep[e.g.,][]{1988Natur.333..523R}, resonant relaxation~\citep[e.g., ][]{RT96,2011PhRvD..84d4024M,2015ApJ...804..128M}, stellar collision\citep[e.g., ][] {1983ApJ...268..565D,1990ApJ...356..483Q,2002A&A...394..345F}, binary effects~\citep[e.g.,][]{2009ApJ...700.1933H, Zhang19, Zhang21}.  Many of these complications are difficult to simulate or include in N-body simulations or analytical methods (for a brief review see~\citep{ZA24}). 

{ Monte Carlo methods provide greater flexibility for incorporating complex physical processes. For instance, the Hénon scheme~\citep[e.g.,][]{1961AnAp...24..369H}, which models relaxation using shell-like particles, has been successfully applied in simulations of globular clusters~\citep[e.g.,][the ``CMC'' code]{2000ApJ...540..969J,2022ApJS..258...22R} and galactic nuclei~\citep[][the ``ME(SSY)'' code]{2002A&A...394..345F}. Other notable approaches include the Princeton scheme~\citep[e.g.,][]{1971ApJ...164..399S}, which allows for the simulation of non-spherical systems~\citep[e.g.,][the ``Raga'' code]{2015MNRAS.446.3150V}, and the Cornell scheme~\citep[e.g.,][]{SM78,1980ApJ...239..685M}, which is particularly advantageous for simulating orbit-related processes across a broad range of energies~\citep[e.g.,][]{ZA24}.}

In a previous study~\citep{ZA24}, we developed a Monte-Carlo method (called \GNC), which was the first to extend the Cornell Monte-Carlo method~\citep{SM78} to include multiple mass components of stellar objects and to simulate the two-body relaxation process by solving the two-dimensional Fokker-Planck (FP) equations.  However, in that work we have ignored the effects of the stellar potential, which has limited its applications to regions where the stellar potential is important or where the mass of the MBH continues to grow.

We therefore make a major update to \GNC~ by integrating the stellar potential, which satisfies the Poisson equation, and also adopt the adiabatic invariant theory to account for the change in orbital energy in response to the slowly varying potential. Importantly, these new updates ensure that the time-evolving solution of the cluster is always self-consistent.  From this new version we can now obtain the evolution of the NSC as the mass of the MBH grows. For clarity, we denote the previous version of \GNC~in~\cite{ZA24} by \GNCp~and the current version of \GNC~ by \GNCc.

So far, the majority of dynamical studies around the MBH have focused on the steady-state solutions for a given mass of the MBH~\citep[e.g.,][]{BW76,1977ApJ...216..883B,SM78,1978ApJ...226.1087C, 2013ApJ...764...52B,2015ApJ...804...52M}. For example, the current rates of tidal disruption of stars in nearby galaxies~\citep[e.g.,][]{1999MNRAS.309..447M,2004ApJ...600..149W,2023ApJ...952..135C} or the event rate of EMRIs~\citep[e.g.,][]{2006ApJ...645L.133H,2006ApJ...645.1152H,2013ApJ...764...52B}.  In reality, MBHs should grow from a seed black hole with a mass of about $10^3-10^4\,\msun$ \citep[e.g.,][]{2020ARA&A..58...27I} and then accrete various surrounding materials, such as gas from falling interstellar medium\citep[e.g.,][]{2010MNRAS.407.1529H}, stellar collisions~\citep[e.g., ][]{1983ApJ...268..565D,1987ApJ...321..199Q,2002A&A...394..345F}, stars via tidal disruption events~\citep[e.g.,][]{1983ApJ...268..565D,2004MNRAS.352..655A,2010MNRAS.405..194F,2012MNRAS.419...57F,2017MNRAS.467.4180S} and swallow other compact objects as they fall into the last stable orbits of the MBH~\citep[e.g.,][]{2022hgwa.bookE..17A}.  Meanwhile, the density profile of the cluster is constantly evolving, and its evolution can change in response to the growth of the MBH masses~\citep[e.g.,][]{1980ApJ...242.1232Y}. 

As a first study, here we employ the updated version of \GNCc~to study the cosmological evolution of the NSCs and the mass of the MBH, assuming that the mass growth of the MBH is purely due to the accretion of stellar objects falling into the loss cone, {including tidal disruption of stars  or 
non-flaring events; i.e. stars being swallowed directly if their orbital pericenter is within $2r_g=2G\bh/c^2$ or, in the case of SBHs, $8r_g$}. \GNCc~gives us properties of the cluster as it evolves, which we can use to make comparisons with those from current observations. Other complications, such as stellar collisions, accretion of gaseous matter and the production of gravitational wave sources are on-going work and will be presented elsewhere.

The paper is organized as follows. In Section 2 we describe the major update made to \GNCc.  In Section 3 we describe the standard tests of Plummer core collapse and the tests of adiabatic takeover of the cluster in response to the mass growth of MBH.  In section~\ref{subsec:fixing_mbh_mass} we first study the evolution of the NSC when the mass of MBH is fixed. We also check the consistency between the results of \GNCp~ and \GNCc, and describe how the evolution of the cluster is changed by including the self-consistent solutions in \GNCc~ rather than the steady-state solution obtained from \GNCp.  In section~\ref{subsec:evolving_mbh_mass} we study the evolution of the cluster and the MBH when the MBH grows by loss-cone accretion from a seed black hole within a Hubble time, and compare the density profile, effect radius, and final mass of the MBH with those from current observations. We also discuss the time-evolving rates of tidal disruption over cosmic time.  The conclusions and discussions are given in section~\ref{sec:conclusion_discussion}.

\section{The method}

The Monte-Carlo method is based on our recently developed \GNCp~code, which calculates the two-body relaxation of multi-mass and multi-type particles by the two-dimensional (energy and angular momentum) FP equations. For more details on the Monte Carlo method see~\citet{ZA24}.  In section~\ref{subsec:units} we introduce the units adopted.  Then we describe the main updates in \GNCc: including the stellar potential (Section~\ref{subsec:including_sp}), adopting the adiabatic invariant theory (Section~\ref{subsec:including_ait}), and updating the diffusion coefficients to include the effects of the stellar potential (Section~\ref{subsec:update_df}).  These new updates ensure that the self-consistent solutions are obtained at all iterations of the simulation. {We also 
discuss the accuracy of the method in presence of some anisotropy of the cluster in Section~\ref{subsec:javg_gx}}; finally, we describe the boundary conditions in \GNCc (Section~\ref{subsec:bd_condition}) and 
how we set the initial conditions of the cluster and generate the initial samples (Section~\ref{subsec:ini_cond}).

\subsection{The units of the simulation and the dimensionless quantities}	
\label{subsec:units}

In the simulation we define a characteristic mass $m_0$ as the unit of mass.  In principle, $m_0$ can be any value. For simplicity, for simulations where the mass of the MBH is fixed, we set $m_0=\bh$.  In other cases, where the mass of the MBH is growing, we set $m_0$ to a fixed value, e.g. $m_0=10^7\,\msun$.

Given $m_0$, for convenience we introduce a characteristic radius $r_0$ as the unit of length, which follows the relation 

\be
r_0=\frac{G m_0}{\sigma_0^2}=3.1~{\rm pc}\times \left(\frac{m_0}{4\times 10^6\msun}\right)^{0.55}.
\label{eq:rh}
\ee

\noindent
This radius $r_0$ can be reduced to the gravitational influence radius of an MBH (defined by $r_h=\bh G/\sigma_0^2$, where $\sigma_0$ is provided by 
$M-\sigma$ relation~\citep{2013ARA&A..51..511K}) by assuming $m_0=\bh$, where $G$ is the gravitational constant. Then the corresponding unit of velocity is $\sigma_0=\sqrt{m_0G/r_0}$, and the unit of number density (mass density) is $n_0=r_0^{-3}$($\rho_0=m_0/r_0^3$).

\subsection{Including stellar potentials}
\label{subsec:including_sp}
\subsubsection{Stellar orbits}

The dynamics of the particles becomes more complicated when the stellar potential is included.  Since the stellar orbits are no longer Keplerian, we have to calculate the periapsis, apoapsis and orbital period numerically\citep{1979ApJ...234.1036C}.

Assuming that the mass density of the cluster as a function of radius $r$ is given by $\rho(r)$, with $r=\infty$ as the reference position, the stellar potentials ($\phi_\star$, for all stellar objects) are given by 

\be\ba
\phi_\star(r)&=4\pi G\left[\frac{1}{r}\int^r_0 \rho(s){s}^2 ds +\int^\infty_r \rho(s)sds\right]\\
&=\frac{M_\star(<r)G}{r}+4\pi G \int^\infty_r \rho(s)sds
\label{eq:potential}
\ea\ee

\noindent
where $M_\star(<r)$ is the enclosed mass within radius $r$. 

The total potential is then given by summing up the part of the stellar objects and those of
MBH, i.e.,

\be
\phi(r)=\phi_\bullet(r)+\phi_\star(r)
\ee

\noindent
where $\phi_\bullet=\bh G/ r$ is the potential of the MBH.
For a given particle, suppose that the velocity and the radial velocity at radius $r$ are given by 
$v$ and $v_r$, respectively, then the specific energy ($E$) of the particle  is defined by 

\be
E=\phi(r)-\frac{1}{2}v^2=\phi(r)-\frac{1}{2}v_r^2-\frac{(jJ_c(E))^2}{2r^2}
\ee

\noindent
where $J_c(E)$ is the maximum angular momentum that corresponds to a circular orbit 
for a given energy $E$, $J$ and $j=J/J_c(E)$ is the angular momentum and the 
dimensionless angular momentum, respectively.


If $E>0$, $r$ oscillates between the pericentre and the apocentre, which correspond to the two roots of the following equation

\be
\frac{1}{2}v^2_r=-E+\phi(r)-\frac{(jJ_c(E))^2}{2r^2}=0
\label{eq:rarp_eq}
\ee

To solve the apocenter $r_a$ and pericenter $r_p$ given  $E$ and $j$,  we need to first obtain the circular angular momentum $J_c(E)$. Given $E>0$, the circular orbit has $r_a=r_p=r_c$. Hence, Equation~\ref{eq:rarp_eq} satisfies~\citep{1979ApJ...234.1036C}

\be
\left.\frac{\pl (J^2)}{\pl r}\right|_{r=r_c}
=\left.\frac{2J^2_c}{r_c}+2\frac{\pl \phi}{\pl r}\right|_{r=r_c}r_c^2=0
\label{eq:circ_orb}
\ee

Combining Equation~\ref{eq:circ_orb} and~\ref{eq:rarp_eq} we can solve $J_c$ and $r_c$,

\be\ba
2[\phi(r_c)&-E]+r_c\left.\frac{\pl \phi}{\pl r}\right|_{r=r_c}=0\\
J_c^2&=-r_c^3\left.\frac{\pl \phi(r)}{\pl r}\right|_{r=r_c}
\label{eq:rcjc}
\ea
\ee

\noindent
After solving $J_c$, we can then solve $r_p$ and $r_a$ for given values of $E$ and $J$ thanks to Equation~\ref{eq:rarp_eq}.

Finally, the orbital period is given by the following integration

\be
P(E,J)=2\int^{r_a}_{r_p}\frac{dr}{v_r}=2\int^{r_a}_{r_p}\frac{dr}{\sqrt{2(\phi(r)-E)-J^2/r^2}}\\
\label{eq:pd}
\ee

\noindent
Note that the period is a function of $E$ and $J$.

\subsubsection{Solving Poisson Equation}
\label{subsec:sp_possion_equaiont}

In \GNCc~we make sure that the stellar potential of the cluster always satisfies the Poisson equation

\be
\nabla \phi_\star(r)=\frac{1}{r^2}\frac{\pl}{\pl r}
\left(r^2\frac{\pl \phi_\star(r)}{\pl r}\right)=4\pi G\rho_\star(r),
\ee

\noindent
for both the initial cluster before the start of the simulation and after each iteration of the simulations. 
In \GNCc~ the simulation of the particles is run for a few fractions of the relaxation time, and then the stellar potential (and the MBH mass, if it is considered to evolve) is updated from the new energy and angular momentum distribution of the particles.  The solution of the stellar potential satisfying the Poisson equation can be found by the iteration method similar to that in~\citet{1979ApJ...234.1036C}.  The details are as follows.

When there are multiple mass components, the number distribution function and the phase distribution function of the $\alpha$-th mass bin can be given by $N_\alpha(E,J)$ and $f_\alpha(E,J)$ in the space of energy $E$ and angular momentum $J$, respectively.  It is convenient to define the following dimensionless distribution 

\be
g_\alpha(x,j)=(2\pi \sigma_0^2)^{3/2}n_0^{-1} f_\alpha(E,J).
\ee

\noindent
where $x=E/\sigma_0^2$ is the dimensionless energy and $j=J/J_c$ is the dimensionless angular momentum. It relates to the number distribution $N_\alpha(x,j)$ by 

\be\ba
g_\alpha(x,j)&=\frac{N_\alpha(x,j)}{2^{3/2}\pi^{1/2} r_0^{3}n_0  j  J_c^{2}  P(x,j)}
\label{eq:g_alphaxj}
\ea\ee

We can define the $j$-averaged distribution according to~\citet{1980ApJ...239..685M}
\be
g_\alpha(x)=\frac{1}{2}\int_0^1 g_\alpha(x,j)j dj
\label{eq:g_alphax}
\ee

Let $\phi_{\star,1}=\phi_{\star,\rm old}$.  The initial condition $\phi_{\star,\rm old}$ is the potential of the input density $\rho_{\rm ini}$. During each iteration of the simulation runs, $\phi_{\star,\rm old}$ is the old potential from the previous step.  Starting from $n=1$, the process is as follows 

\begin{enumerate} 

\item We count the number of particles in the $\alpha$-th mass bin and get $N_\alpha(E,J)$ in bins of $E$ and $J$.  

\item We obtain the dimensionless function $g_\alpha(x)$ from $N_\alpha(E,J)$ of the $\alpha$-th mass bin by the equation~\ref{eq:g_alphaxj} and~\ref{eq:g_alphax} according to the function of $J_c(E)$ and the period $P(E,J)$ solved by the equation~\ref{eq:rcjc} and~\ref{eq:pd}.  

\item Calculate the new mass density distribution $\rho_{\star,n+1}(r)$ given the updated distribution function $g_\alpha(x)$ from above, i.e,

\be\ba
\rho_{\star,n+1}(r)&=\frac{2^{1/2}}{\pi^{1/2}} n_0\sum_\alpha m_\alpha \\
&\times \int^{ \phi_{\star,n}( r)}_0
g_\alpha(x)\sqrt{2( \phi_{\star,n}( r)-x)}dx
\label{eq:rho}
\ea
\ee

\noindent
where $m_\alpha$ is the mass of the $\alpha$-th mass bin.

\item We can then replace $\rho_{\star,n+1}$ into the Equation~\ref{eq:potential}
to obtain the updated potential $ \phi_{\star,n+1}$. 

\item We calculate the convergence of potential by 
$${\rm C}_\phi
=\left\langle \left|\log \frac{ \phi_{\star,n+1}(r)}{ \phi_{\star,n}(r)}\right| \right\rangle$$ where $\langle \cdot \rangle$ is averaging over all radius across the cluster.
{If ${\rm C}_\phi<10^{-3}\sim10^{-4}$, we assume that the potential has converged}. Otherwise set $\phi_{\star,n+1}\rightarrow\phi_{\star,n}$ and go back to step 2.  

\end{enumerate} 

Usually the potential converges within $5-10$ iterations. We then obtain the new self-consistent stellar potential $\phi_{\star,\rm new}$ according to the updated samples.

We find that at $x_T\simeq \phi(r_T)$, where $\phi(r_T)=\bh/r_T$ (or $\phi(0)$ if $\bh=0$), the distribution function of $g(x)$ can change rapidly, mainly due to the rapid decrease of the volume function of the phase~\citep{1995ApJ...440..554Q}, which is given by

\be
b(x)=2^{3/2}\int^{\phi^{-1}(x)}_0r^2\sqrt{\phi(r)-x}dr.
\ee

In some cases (e.g. when $\bh$ is small compared to the total mass of the cluster), when uniform bins of dimensionless energies $x$ are adopted, such a rapid decrease of $g(x)$ can lead to a divergence of $C_\phi$ in the iterations of finding a self-consistent solution of $\phi_\star$.

To obtain a stable convergence of $C_\phi$, we set bins of energies such that the coverage of $g(x)$ is denser in regions $x \sim \phi(r_T)$.  We find that the derivative of $J_c(x)$ provides a very smooth and natural way to set the energy bins (for a typical $J_c(x)$ see the upper left panel of Figure~\ref{fig:rcjc}).  We set the bin size at energy $x$, i.e. $h(x)$, such that

\be
h(x)\propto \left(\frac{d\ln J_c(x)}{d\ln x}\right)^{-1}.
\ee

Then $h(x)$ is constant if $x\gg \phi(r_T)$ or $x\ll \phi(r_T)$ and becomes 
smaller near $x\sim \phi(r_T)$.  We note that {in most cases} the use of $h(x)$ above  
leads to a convergence of $C_\phi$.  {However, when $\bh=0$ and the density profile resembles a core-like structure (e.g., the Plummer model), we find that $h(x)\rightarrow0$ as $x\rightarrow \phi(0)$. This behavior can cause large fluctuations in $g(x)$ because the number of samples within these bins becomes very small. In such cases, we impose an additional requirement for a minimum bin size, defined as \( h(x)=(x_{\rm max}-x_{\rm min})/N_{\rm res} \), where \( N_{\rm res} \) is a constant, typically in the range of $200\sim500$. Here, \( x_{\rm max} \) and \( x_{\rm min} \) represent the maximum and minimum dimensionless energy at the current time in the simulation.}.

\subsection{Including adiabatic invariant theory}
\label{subsec:including_ait}

As the stellar potential and the mass of the MBH evolve, the orbital energy of individual stellar objects will change in response to the change in the potential around them, even if other relaxations, such as the two-body relaxation process, are ignored. According to the adiabatic invariant theory, the orbital energy is no longer conserved, but the orbital angular momentum ($J$) and the radial action ($Q$) are invariants in a slowly varying spherical potential~\citep{1973LyndenBell,1980ApJ...242.1232Y}.  The radial action is given by

\be
Q=2\int_{r_p}^{r_a} v_r dr={\rm constant}
\ee

\noindent
For each particle, $Q$ is conserved before and after the updates of potential. Thus
\citep{1973LyndenBell,1980ApJ...242.1232Y}

\be\ba
Q&=2\int_{r_{p,\rm new}}^{r_{a,\rm new}} \sqrt{2\left(\phi_{\rm new}+\frac{M_{\bullet,\rm new}}{r}
-E_{\rm new}\right)-\frac{J^2}{r^2}} dr\\
&=2\int_{r_{p,\rm old}}^{r_{a,\rm old}} \sqrt{2\left(\phi_{\rm old}+\frac{M_{\bullet,\rm old}}{r}
-E_{\rm old}\right)-\frac{J^2}{r^2}} dr
\label{eq:q_ad}
\ea\ee

Here $r_{a,\rm old}$, $r_{p,\rm old}$ ($r_{a,\rm new}$, $r_{p,\rm new}$) are the apoapsis and periapsis of each particle before (after) the potential update, respectively.      $M_{\bullet,\rm old}$ is the mass of the MBH in the previous iteration.  $M_{\bullet,\rm new}$ is the updated mass of the MBH in the current iteration of the simulation (e.g. due to loss-cone accretion of stars and SBHs). We then added the energy drift $E_{\rm new}\rightarrow E_{\rm old}+\delta E$ for each of the particles after each iteration of the simulation.  The invariant of the radial action then shifts the previous energy ($E_{\rm old}$) to a new one ($E_{\rm new}$). 

The $E_{\rm new}$ of individual particles must be solved explicitly from the equation~\ref{eq:q_ad}, which is quite tedious and time-consuming if done for a large number of particles (usually $>10^5$).  Therefore, a more convenient but slightly less accurate method can be given by~\citep{1979ApJ...234.1036C}

\be\ba
&\delta E=E_{\rm new}-E_{\rm old}\simeq \left\langle\frac{dE}{dt}\right\rangle \delta t\\
=&\frac{2}{P}\int^{r_{a,\rm old}}_{r_{p,\rm old}} \frac{\pl \phi(r,t)}{\pl t}\delta t\frac{dr}{v_r}\\
=& \frac{2}{P}\int^{r_{a,\rm old}}_{r_{p,\rm old}} \left(\phi_{\star,\rm new} 
-\phi_{\star,\rm old} +\frac{M_{\bullet,\rm new}-M_{\bullet,\rm old} }{r}\right)\frac{dr}{v_r}.
\label{eq:de}
\ea\ee

We find that the difference between $E_{\rm new}$ from Equation~\ref{eq:de} and Equation~\ref{eq:q_ad} is  about $1\%$.

\subsection{The diffusion coefficients when the stellar potential is taken into account}
\label{subsec:update_df}

In \GNCc, the diffusion coefficients of each component need to be updated 
in order to consider the effects of stellar potential.
Similar to~\citet{1979ApJ...234.1036C}, for the  $\alpha$-th component,
we define the following dimensionless function that are useful for the integration of the
diffuse coefficients

\be\ba
 F^\alpha_0(x)&=\int^1_{-\infty} g_\alpha(sx) ds \\
 F^\alpha_1(x,  r)&=\int^{ \phi( r)/x}_{1} g_\alpha(sx)\sqrt{\frac{ \phi( r)/x-s}{ \phi( r)/x-1}} ds\\
 F^\alpha_3(x, r)&=\int^{ \phi( r)/x}_{1} g_\alpha(sx)\left(\frac{ \phi( r)/x-s}{ \phi( r)/x-1}\right)^{3/2} ds.
\label{eq:ffunc_seris}
\ea\ee

Then the updated diffusion coefficients are given by~\citet{1979ApJ...234.1036C}

\be
\ba
\frac{D_E}{E}&=\frac{2\mathcal{C}}{ P}\int^{ r_p}_{ r_a} 
\sum_\beta \left(m_\beta m_\alpha  F_1^\beta -m_\beta^2  F_0^\beta\right) \frac{d r}{ v_r}
\ea\ee
\be\ba
\frac{D_{EE}}{E^2}&=\frac{8\mathcal{C}}{3 P}\int^{ r_p}_{ r_a} \sum_\beta m_\beta^2 
\left(\frac{ \phi}{x}-1\right)\left( F_3^\beta+ F_0^\beta\right) 
\frac{d r}{ v_r}
\ea\ee
\be\ba
\frac{D_J}{J_c}&=\frac{\mathcal{C}}{ P}\int^{ r_p}_{ r_a} \left[ \frac{x  r^2 }{j{ J}_c^2} 
\sum_\beta m_\beta^2 \left( F_1^\beta-\frac{1}{3} F_3^\beta+\frac{2}{3} F_0^\beta\right)\right.\\
&\left. 
-\frac{j}{ \phi/x-1}\sum_\beta m_\beta(m_\alpha+m_\beta)  F_1^\beta \right]
\frac{d r}{ v_r}
\ea\ee
\be\ba
\frac{D_{JJ}}{J^2_c}&=\frac{2\mathcal{C}}{ P}\int^{ r_a}_{ r_p}\left[\frac{ j^2}{3} 
\frac{\sum_\beta m_\beta^2  F_3^\beta}{\phi/x-1}\right.\\
&+\left(\frac{ r^2 x}{ J_c^2}
-\frac{j^2}{2(\phi/x-1)}\right) 
\sum_\beta m_\beta^2 \left( F_1^\beta-\frac{1}{3} F_3^\beta\right)\\
&\left.+\frac{2}{3}  \frac{ r^2 x}{{ J}_c^2} \sum_\beta m_\beta^2  F_0^\beta\right] 
\frac{d r}{ v_r}
\ea\ee
\be\ba
\frac{D_{EJ}}{EJ_c}&=-\frac{4\mathcal{C} j}{3 P}\int^{ r_a}_{ r_p} 
\sum_\beta m_\beta^2 ( F_3^\beta+ F_0^\beta)\frac{d r}{ v_r}
\ea\ee
where 
\be\ba
 v_r&=\sqrt{2 \phi( r)-2x-\frac{j^2  J_c^2 }{ r^2}}\\
 P&=2\int^{ r_a}_{ r_p}\frac{d r}{ v_r}\\
\mathcal{C}&=n_0(2\pi \sigma_0^2)^{-3/2} \kappa,~~\kappa=16\pi^2 G^2 \ln \Lambda
\ea\ee
where $\Lambda=0.4 N_{\rm tot}$, $N_{\rm tot}=\sum_\alpha N_\alpha$ 
is the total number of stellar objects in the whole cluster, and $N_\alpha$ is the number of 
objects in the $\alpha$-th mass bin.

Similar to \GNCp, these diffusion coefficients are calculated in tabular before the beginning of each iteration of the simulation.
Although there is no auxiliary function similar to those in \GNCp,  
we can linearly interpolate the function $g_\alpha(x)$ such that
 integration of Equation~\ref{eq:ffunc_seris} can be done quite quickly. For the 
integration of diffusion coefficients, 
we use the DOPRI5 based on the explicit fifth(fourth)-order Runge Kutta method~\citep{DP80, Hairer93}.
The tabular diffusion coefficients for each mass bin can be calculated in parallel and completed within several seconds on a standard desktop computer with 24 threads.
\subsection{The accuracy of the results in presence of some anisotropy of the cluster}
\label{subsec:javg_gx}
{Since we use the $j$-averaged value of the distribution for a given energy \(x\), i.e., \(g(x)\) in Equation~\ref{eq:g_alphax}, the estimation of diffusion coefficients and density in Equation~\ref{eq:rho} may become less accurate in regions with strong anisotropy. It is important to note, however, that \(g(x,j)\) can be an arbitrary function of \(j\). As a result, \(g(x)\), estimated using Equation~\ref{eq:g_alphax}, is generally not equivalent to the distribution assumed under isotropy.}

{According to~\citet{1979ApJ...234.1036C}, a more general estimation of the diffusion coefficients in an anisotropic cluster should adopt
\be
g_\alpha(x,r)=\int_0^1\frac{g_\alpha(x,j_0j')j'dj'}{\sqrt{1-j'^2}},
\label{eq:gxr}
\ee
in Equation~\ref{eq:ffunc_seris}, where \( j_0=\frac{r}{J_c}\sqrt{2(\phi-x)} \) is the maximum dimensionless angular momentum for particles with energy \(x\) located at position \(r\). }

{The density profile of the cluster is then given by}
\be
\rho=2^{1/2}\pi^{-1/2}n_0 \sum_\alpha m_\alpha \int_0^{\phi(r)}g_\alpha(x,r)\sqrt{2(\phi-x)}dx.
\label{eq:rho_gxr}
\ee

{
We then perform additional simulations to compare the results obtained by adopting \(g(x,r)\) from Equation~\ref{eq:gxr} and \(g(x)\) from Equation~\ref{eq:g_alphax} for a given cluster that includes loss cone effects. The differences are generally small across most regions of the cluster. However, noticeable differences appear in the innermost region near the MBH, where strong anisotropy exists due to particles predominantly following near-circular orbits. This region corresponds to energies of \(x \gtrsim 10^5\) or distances of \(r \lesssim 10^{-5}\) pc for a Milky-Way-like NSC, which lies outside the primary regions of interest in this study. 
For the simulations conducted in this study, no significant differences are observed between the two methods in terms of the evolution of density profiles and tidal disruption rates of stars. These results suggest that the NSCs examined here do not exhibit strong anisotropy overall, and thus using the \(j\)-averaged \(g(x)\) provides sufficient accuracy for simulating the dynamics in most parts of the cluster.
}

{In principle, we can use \(g(x,r)\) for our simulations, allowing for a more general treatment of anisotropic systems. However, for the simulations performed in this study, we use the \(j\)-averaged \(g(x)\) primarily because it is approximately $4\sim10$ times faster than the version adopting \(g(x,r)\).
}

\subsection{Boundary conditions}
\label{subsec:bd_condition}

In the previous version, a Dirichlet boundary condition ($g(x_b)=1$) was adopted at some low-energy boundary ($x_b$) in order to normalise the weightings of particles, thereby ensuring that the density of particles at the influence radius of the MBH ($r_h$), defined by $r_h=\bh/\sigma_0^2$ following the $M-\sigma$ relation, was approximately equal to a given value of $n_h$. 
This is due to the fact that \GNCp~does not incorporate the stellar potential in a self-consistent manner. 
Such a normalisation can be used to align the density distribution from the inner regions with an isothermal density distribution at the outer regions.

In the updated version, \GNCc, the entire star cluster can be simulated by including the stellar potential, eliminating the need for patch solutions.
The boundary condition at the outer parts of the cluster is effectively an evaporating boundary condition. It is assumed that particles are ejected from the system if their energy is very close to zero, that is to say, if $x<x_{\rm min}$, where $x_{\rm min}$ corresponds to the energy of a circular orbit with a radius of $r_{\rm max}$ approximately equal to 1.5 kpc.

The remaining boundary conditions are in accordance with those of the aforementioned \GNCp.
For instance, particles are removed if they traverse the high-energy boundary 
at $x=x_{\rm max}=10^5\sim 10^6$, which corresponds to the circular orbit of radius $8\sim 80r_g$ 
for a Milky Way MBH. {$x_{\rm max}$ can be higher if necessary.}
The same boundary condition for angular momentum is adopted as in \GNCp. At high angular momentum, no flux crosses the boundary $j=1$. If the loss cone is considered, particles are removed when they reach orbital pericentre and within the loss cone of angular momentum $j_{\rm lc}$, i.e. $j<j_{\rm lc}$.

\subsection{The initial distribution functions of the cluster}
\label{subsec:ini_cond}

Initially, we need to generate a large number of particles with a distribution of 
orbits (specified by pairs of $E$ and $J$) that corresponds to 
some given initial density profile $\rho_{\rm ini}(r)$. They are the 
initial Monte-Carlo samples of the simulation. 

The corresponding potential is then $\phi_{\rm ini}=\phi_\bullet+\phi_{\star,\rm ini}$, where $\phi_{\star,\rm ini}$
is given by Equation~\ref{eq:potential} given $\rho=\rho_{\rm ini}$. 
Assuming an isotropic distribution of angular 
momentum, we can calculate the phase-space energy distribution 
function $\bar f_{\rm ini}(E)$~\citep{Binney87}
\be\ba
\bar f_{\rm ini}(E)&=\frac{1}{2^{3/2}\pi^{2}}\left[\int^{E}_{E_{\rm min}}
\frac{\pl^2 \rho_{\rm ini}}{\pl \phi^2}
\frac{d \phi}{\sqrt{E- \phi}}\right.\\
&\left.+\frac{1}{\sqrt{E}}\left(\frac{d  \rho_{\rm ini}}{d  \phi}\right)_{\phi=E_{\rm min}}\right],
\label{eq:fe}
\ea\ee
where $E_{\rm min}=x_{\rm min}\sigma^2_0$ 
is the minimum value of the energy determined by the finite outer edge of the cluster (See Section~\ref{subsec:bd_condition}).

The orbits of the particles are then fully
determined by specifying the number distribution of 
energies, i.e., $N_{\rm ini}(E)$, which is given by the following integration 
\be
N_{\rm ini}(E)=16\sqrt{2} \pi^{2}  \bar f_{\rm ini}(E)
 \int^{\phi^{-1}_{\rm ini}(E)}_{0} r^2 \sqrt{\phi_{\rm ini}(r) -E} d r
\ee
\noindent
From this distribution, a group of initial samples at $E$ can be generated according to $N_{\rm ini}dE$ given energy bins of $E$ and $E+dE$ by 
Monte Carlo methods~\footnote{We notice that an alternative method of initialization is first generating Monte Carlo samples in space of 
velocity and position according to Equation 22 of~\citep{2012ApJ...745...83A} 
(originally presented in~\citet{2005PhRvL..95h1102S}), and then convert them back to the space of energy and angular momentum.}.

Subsequently, the procedures outlined in Section~\ref{subsec:sp_possion_equaiont}  are employed to derive the self-consistent solution for the potential and density. 
It should be noted that the self-consistent solution of the density and potential may differ slightly from the initial values, due to the Monte Carlo fluctuations.

It can be demonstrated that Equation~\ref{eq:fe} does not necessarily guarantee a positive solution for the initial energy function, $\bar f_{\rm ini}(E)$, in the presence of a MBH at the centre. 
In such instances, the self-consistent solution of the initial cluster may fail to converge with the intended distribution of density.
In order to circumvent this issue, this study exclusively considers scenarios where a positive solution is present across the relevant energy range. 

\section{Model Tests}

\begin{figure*}
	\center
	\includegraphics[scale=0.5]{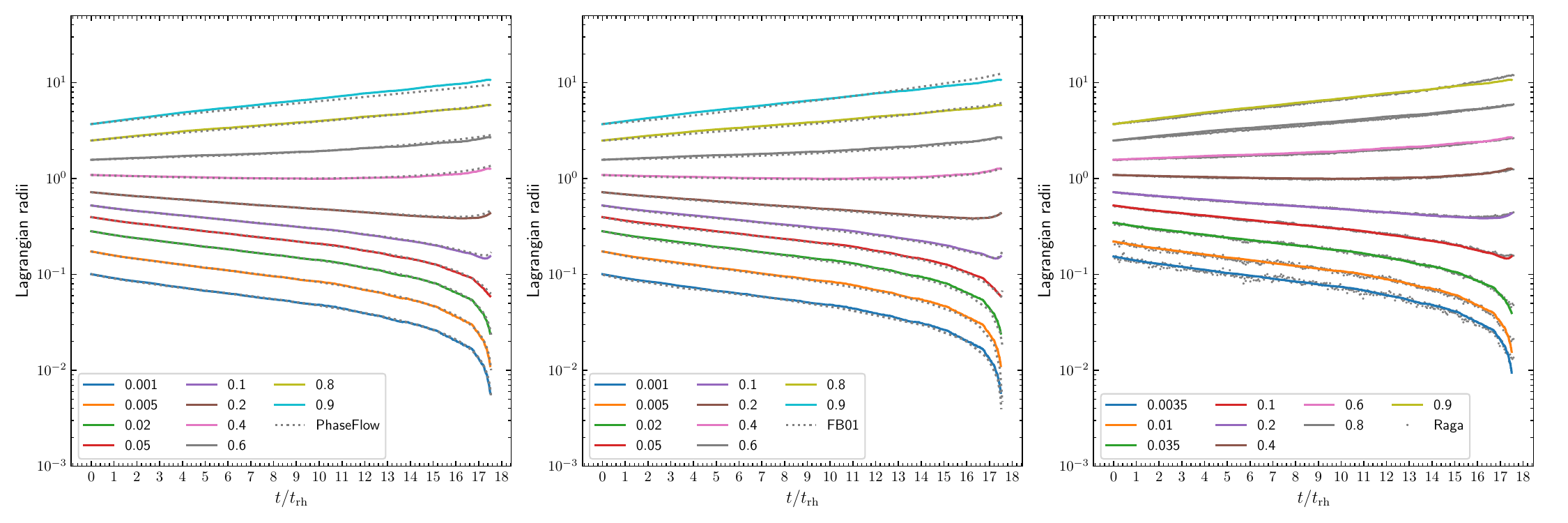}
	\caption{
 The evolution of the Lagrangian radii during the
core-collapse  of a Plummer cluster with equal mass of stars. 
The colored lines are from \GNCc~and those gray dotted lines or circles correspond to the results 
using the PhaseFlow code~\citep[][left panel]{2015MNRAS.446.3150V,2017ApJ...848...10V}, 
ME(SSE) code~\citep[][middle panel]{2001A&A...375..711F}, and Raga ~\citep[][right panel, kindly provided by Eugene Vasiliev]{2015MNRAS.446.3150V}. 
The legends show the corresponding fraction of masses within the Lagrangian radius. 
Here the Coulomb factor (${\rm lg}\Lambda$) in PhaseFlow, ME(SSE)
and Raga are adjusted to make a comparison with \GNCc. 
}
\label{fig:lag}
\end{figure*}

\begin{figure*}
	\center
	\includegraphics[scale=0.7]{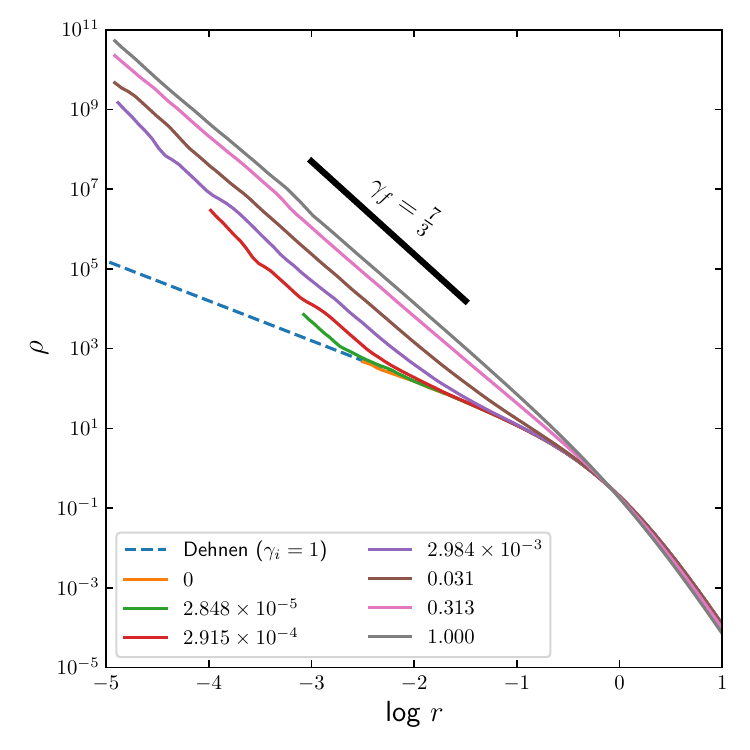}
	\includegraphics[scale=0.7]{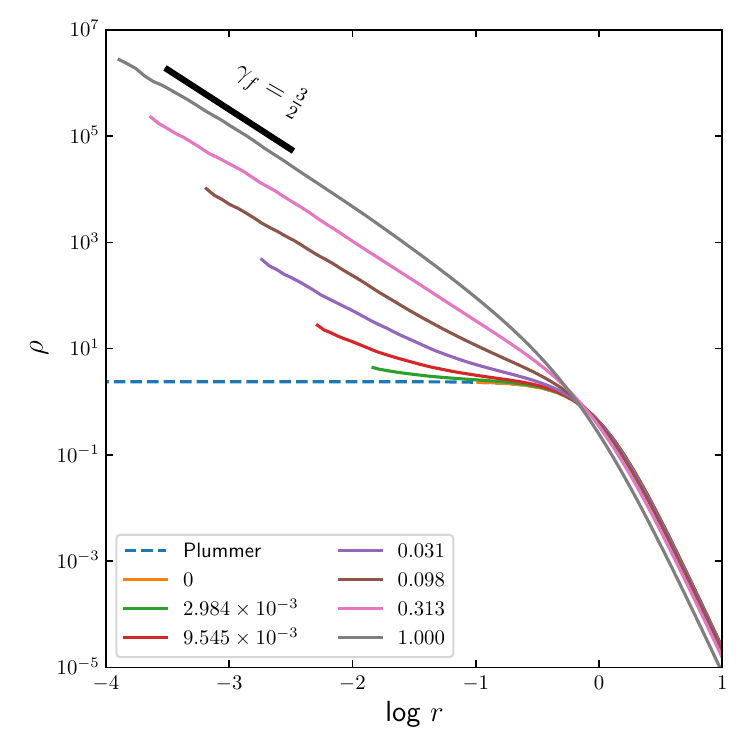}
	\caption{The adiabatic adoption of the density profile due to the mass growth of the MBH. 
	The colored solid lines show the density distribution of stars when
     the mass of the MBH grows up to some fraction 
     of the initial mass of the cluster, i.e., $\bh/M_{\rm cl}(0)$, as shown in the legend. 
     The blue dashed lines in both panels show the initial density distribution {of the 
	 model}.
	Note that in these two tests the two-body relaxation and the effects of loss cone are turned off. The radius 
    $r$ is in unit of $r_0$ and $\rho$ in unit of $\rho_0$.
	}
\label{fig:adiab}
\end{figure*}

It is of great importance to conduct a series of tests on \GNCc~in order to ascertain 
that it is capable of producing satisfactory results. In Section~\ref{subsec:core_collapse}, we 
demonstrate that the implementation of \GNCc~can accurately reproduce the core collapse 
phenomenon observed in Plummer clusters. The time of core collapse is found to be consistent with that 
observed in numerous previous studies. Secondly, we demonstrate in Section~\ref{subsec:adb_theory} that 
it can also reproduce the expected self-adoption of the density profile, provided that only the mass 
growth of the central MBH is considered.

\subsection{Plummer Core-collapse test of \GNCc}
\label{subsec:core_collapse}

The first test examines the core collapse of an isolated star cluster composed of equal-mass stars, which initially follows the density profile of the Plummer model~\citep[e.g.,][]{1979ApJ...234.1036C,1961AnAp...24..369H,1980ApJ...239..685M,2015MNRAS.446.3150V}. In this scenario, there is no massive black hole (MBH) at the center, and the cluster collapses solely due to its own gravitational instability. The initial Plummer density profile is described by~\citep{1911MNRAS..71..460P}.

\be
\rho_p(r)=\frac{3M_{\rm cl}r_a^2}{4\pi (r^2+r_a^2)^{5/2}},
\label{eq:plummer}
\ee
where $r_a$ is a characteristic radius (which is also the effective radius of the cluster) and
$M_{\rm cl}$ is the initial stellar mass of the cluster. {We conduct tests of Plummer core-collapse using \GNCc. The simulation intervals are set to one-tenth of the two-body relaxation timescale in the innermost region of the cluster, which is found to be sufficiently short to ensure the convergence of the results.}

The characteristics of core collapse are typically illustrated by the evolution of the Lagrangian radii, which enclose specific fractions of the total mass of the cluster. Figure~\ref{fig:lag} presents the evolution of the Lagrangian radii obtained from \GNCc~and compares them with those derived 
from other {methods}, such as PhaseFlow~\citep{2015MNRAS.446.3150V,2017ApJ...848...10V} {{which is a one-dimensional analytical Fokker-Planck method 
that can deal with multiple mass components}}, ME(SSY)~\citep{2001A&A...375..711F}, {{
a Monte-Carlo method adopts H\'{e}non's scheme}}, and Raga~\citep{2015MNRAS.446.3150V}, {{ 
a Monte-Carlo method adopting Princeton's scheme}}. 
All these methods show strong consistency with $N$-body numerical simulations of Plummer core collapse. The horizontal axes represent the simulation time in units of $t_{\rm rh}$, the half-mass relaxation time~\citep{1971ApJ...164..399S}

\be
t_{\rm rh}=\frac{0.06r_{\rm hf}^{3/2}M_{\rm cl}^{1/2}}{m_\star G^{1/2}{\rm lg} \Lambda},
\ee

\noindent
where $r_{\rm hf} = 1.3048r_a$ is the half-mass radius, ${\rm lg} \Lambda \simeq {\rm lg} (0.4N)$, $N = M_{\rm cl}/m_\star$ is the total number of stars, and $m_\star = 1\,\msun$ is the mass of each star.

Some systematic variations in $t_{\rm rh}$ may be introduced because the previously mentioned methods adopt slightly different values of the Coulomb logarithm ${\rm lg} \Lambda$. To enable a reasonable comparison with our method, we have adjusted the values of ${\rm lg} \Lambda$ for all three methods. The evolution of the Lagrangian radii from \GNCc~shows strong consistency with any of these three methods.

The core-collapse time in Figure~\ref{fig:lag} is approximately $\sim 17.6t_{\rm rh}$, which is consistent with values found in the literature, i.e., $14 \sim 18t_{\rm rh}$~\citep[e.g.,][]{1961AnAp...24..369H,1979ApJ...234.1036C,1980ApJ...239..685M,2015MNRAS.446.3150V,2001A&A...375..711F,1995PASJ...47..561T}. In this specific run, \GNCc~utilized approximately 500k particles. It should be noted that Monte Carlo errors due to the limited number of particles, as well as variations in the selected number of bins in energy $E$ and radius $r$, may slightly affect $t_{\rm hf}$. By varying the number of bins in energy and radius in the range of 72–120, and starting with different random seeds for the Monte Carlo simulations, we find that the time of collapse typically occurs around $17.3 \pm 0.5t_{\rm rh}$.

\subsection{Adiabatic evolution of the cluster due to the mass growth of the MBH}
\label{subsec:adb_theory}

For a cluster with an initial density distribution following a power-law, i.e., $\rho_{\rm ini} \propto r^{-\gamma_i}$, and considering only the adiabatic adjustment of particle energies due to the mass growth of a central massive black hole (MBH), while neglecting two-body relaxation (as discussed in Section~\ref{subsec:including_ait}), it is expected that the final slope index of the density profile, when the MBH grows to a mass comparable to that of the initial cluster mass, will follow the relationship described by ~\citep{1995ApJ...440..554Q}.

\be
\gamma_f=\frac{3}{2}+n\frac{2-\gamma_i}{4-\gamma_i}.
\ee

\noindent
where $n$ is the slope of the energy distribution $f(E)$ near $E = \phi(0)$, such that $f(E) \propto (E - \phi(0))^{-n}$, and $\phi(0)$ represents the potential at the center of the cluster (in the absence of an MBH).

To test the adiabatic adjustment of the density, we perform simulations using \GNCc~and adopt two different initial density models. The first model is the Plummer model, as described in Equation~\ref{eq:plummer}, which has $\gamma_i = 0$. The second model is the Dehnen model~\citep{1993MNRAS.265..250D}:

\be
\rho_d(r)=\frac{(3-\gamma) M_{\rm cl}}{4\pi}
\frac{{r}_a}{ r^\gamma(  r+{  r}_a)^{4-\gamma}},
\label{eq:dehnen}
\ee

\noindent
where $r_a$ is a characteristic radius, $M_{\rm cl}$ is the initial stellar mass of the cluster, 
and $\gamma$ is a slope index in the inner regions of the cluster. Here, we test the case where $\gamma_i = \gamma = 1$.

{In these tests, we  disable two-body relaxation and loss cone effects. 
We {first obtain a self-consistent cluster without a MBH, we then
place a MBH at the center and} 
continuously increase the mass of the MBH in a logarithmic fashion,
from a fraction of {$10^{-5}$} of 
the cluster's mass to $100\%$ in 100 iterations.
In each of them we derive a self-consistent density profile of the cluster using 
the method described in Section~\ref{subsec:including_sp}. 
The energy of each particle is adjusted according to
the radial action invariance, as described in Section~\ref{subsec:including_ait}.} 
The results from \GNCc~are shown in Figure~\ref{fig:adiab}. When the mass of the MBH increases to approximately that of the cluster, i.e., $\bh \simeq M_{\rm cl}$, the theoretical expectation for the slope index is $\gamma_f = 7/3$ (where $n = 5/2$) for the Dehnen model with $\gamma_i = 1$~\citep{1995ApJ...440..554Q}, and $\gamma_f = 3/2$ (where $n = 0$) for the Plummer model. The final density slopes obtained for both the Dehnen and Plummer models from \GNCc~are in strong agreement with these theoretical expectations.

\section{Self-consistent solution of evolving nuclear star clusters}
\label{sec:self_consistent_solution}

\begin{table*}
	\caption{Models}
		\centering
	\begin{tabular}{|c|c|c|c|c|c|c|c|c|c|}\hline
	 Name & $M_{\rm cl}^a$ ($\msun$) & $r_a^a$ &$r_{\rm eff,i}^b$ & $r_{\rm eff,f}^c$ & $r_{\rm h,f}^c$
	  & $\bh^{d}$ ($\msun$) & Components$^{e}$ 
	  & $M_{\bullet,f}$ ($\msun$)$^f$ & $R_{td,f} ({\rm yr}^{-1})^g$\\
	\hline
	M1     &  \multirow{3}{*}{$4\times 10^7$} & $2.17$ & $3.9$ & $5.9$ & $3.1$
	&\multirow{3}{*}{$4\times 10^6$, Fix}& stars ($1\msun$)    
		& \multirow{3}{*}{-}& $1.3\times 10^{-4}$\\
	\cline{8-8}
	M2    &    & $2.17$ & $3.9$ & $6.1$ & $3.5$ &
	  & \multirow{18}{*}{\begin{tabular}{c} stars~($1\msun$) 
		\\SBHs~($10\,\msun$)\\ $f_\bullet=0.001$\end{tabular}}     & & $7.4\times 10^{-5}$\\
	{M2\_2}    & &  $2.91$ & $3.9$   &  $5.5$ & $3.3$
		&    & & & $9.6\times10^{-5}$  \\
	\cline{1-7}\cline{9-10}
	M2G91   & \multirow{3}{*}{$10^{9}$}  & $2.56$  & $4.6$ & $5.2$& $1.63$ &\multirow{13}{*}{$10^4$, Growth} &   
		&  $5.2\times 10^7$ & $1.6\times 10^{-3}$\\
	{M2G91\_2} & & $3.4$ & $4.6$ & $5.1$ & $1.7$ & & & $6.3\times10^7$ & $1.8\times 10^{-3}$\\		
	M2G92   &  & $10.26$ & $18.3$  & $18.6$ & $1.48$& &    & $5.3\times 10^6$ & $6.7\times 10^{-4}$ \\
	\cline{1-6}\cline{9-10}	
	M2G81   &  \multirow{5}{*}{$4\times 10^7$}  & $0.5$  & $0.9$ &  $3.0$    &  $5.9$ &   &   &$9.0\times 10^6$& $1.6\times 10^{-4}$ \\
	M2G82   &    & $1.5$  & $2.7$   &  $4.6$  &  $2.7$ &      & &$4.1\times 10^6$& $1.5\times 10^{-4}$  \\
	M2G83   &    & $2.3$  &  $4.1$   &  $5.4$  &  $2.0$ &       &  &$2.2\times 10^6$& $1.3\times 10^{-4}$ \\
	{M2G83\_2}   &    &$3.2$  &  $4.1$  &  $5.9$  &$2.5$  &       &  & $2.8\times10^6$ & $1.5\times10^{-4}$\\
	M2G84   &    & $4.1$  & $7.4$ &   $7.9$  &  $1.1$ &      &  &$2.0\times 10^5$ &  $4.0\times 10^{-5}$ \\
	\cline{1-6}\cline{9-10}
	M2G71      &   \multirow{5}{*}{$2\times 10^{6}$}  & $0.39$  & $0.7$  &  $5.5$ & $3.3$     &   & &$1.6\times 10^5$ &$1.5\times 10^{-6}$  \\
	{M2G71\_2}   &    &  $0.55$ &  $0.7$   &  $5.8$ & $3.6$     &   &  &  $1.6\times10^5$ &  $1.3\times10^{-6}$  \\
	M2G72   &    & $0.72$   & $1.3$&  $5.3$  &  $1.9$ &     & &$6.8\times 10^4$ &$1.2\times 10^{-6}$  \\
	M2G73   &    & $1.4$  & $2.6$&  $5.7$   &  $1.5$ &     &  &$3.6\times 10^4$ & $1.1\times 10^{-6}$  	\\
	M2G74   &    & $2.9$  & $4.7$ &  $7.3$   &  $1.4$ &     &  &$2.1\times 10^4$ & $4.8\times 10^{-7}$  					\\
	\cline{1-7}\cline{9-10} 
	M2G82m   &  $4\times 10^7$  & $1.5$  & $2.7$   &  $4.4$  &  $2.5$ &   \multirow{2}{*}{$10^3$, Growth}    & &$4.1\times 10^6$& $2.0\times 10^{-4}$  \\
	M2G71m   &  $2\times 10^6$  & $0.36$  & $0.7$   &  $5.1$  &  $2.6$ &     & &$1.4\times 10^5$& $3.6\times 10^{-6}$  \\
	\hline
	\end{tabular}
	%
		\tablecomments{		
$^{a}$. $M_{\rm cl}$ and $r_a$ (in units of pc) are the initial mass of the cluster and the characteristic radius of Dehnen's model (Equation~\ref{eq:dehnen}), respectively. {Models with 
suffix ``\_2'' adopt $\gamma=1.5$, those without any suffix have $\gamma=1$};\\
$^{b}$. $r_{\rm eff,i}$ (in units of pc) is the initial effective radius of the cluster;\\
$^{c}$. $r_{\rm eff,f}$ and $r_{\rm h,f}$ (both in units of pc) represent the effective radius and the influence radius of the MBH, respectively, at one relaxation time $T_{\rm rlx}$ for M1, M2 and M2\_2, or at $12$ Gyr for other models;\\
$^d$. The initial mass of the MBH. ``Fix'' indicates that the MBH mass is fixed to the initial value; ``Growth'' indicates that the MBH can grow due to loss-cone accretion (tidal disruption of stars or the direct swallowing of stars falling into event horizon or SBHs with $r_p \leq 8r_g$); \\
$^e$. $f_\bullet$ is the number fraction of SBHs relative to the total number of stars; \\
$^f$. The MBH mass at $12$ Gyr as calculated by \GNCc, for models that consider the mass growth of the MBH;\\
$^g$. The rate of tidal disruption at one relaxation time $T_{\rm rlx}$ for M1, M2 and M2\_2, or at $12$ Gyr for other models.\\
	}	%
	\label{tab:model}
	%
	\end{table*}

\begin{figure*}
	\center
	\includegraphics[scale=0.9]{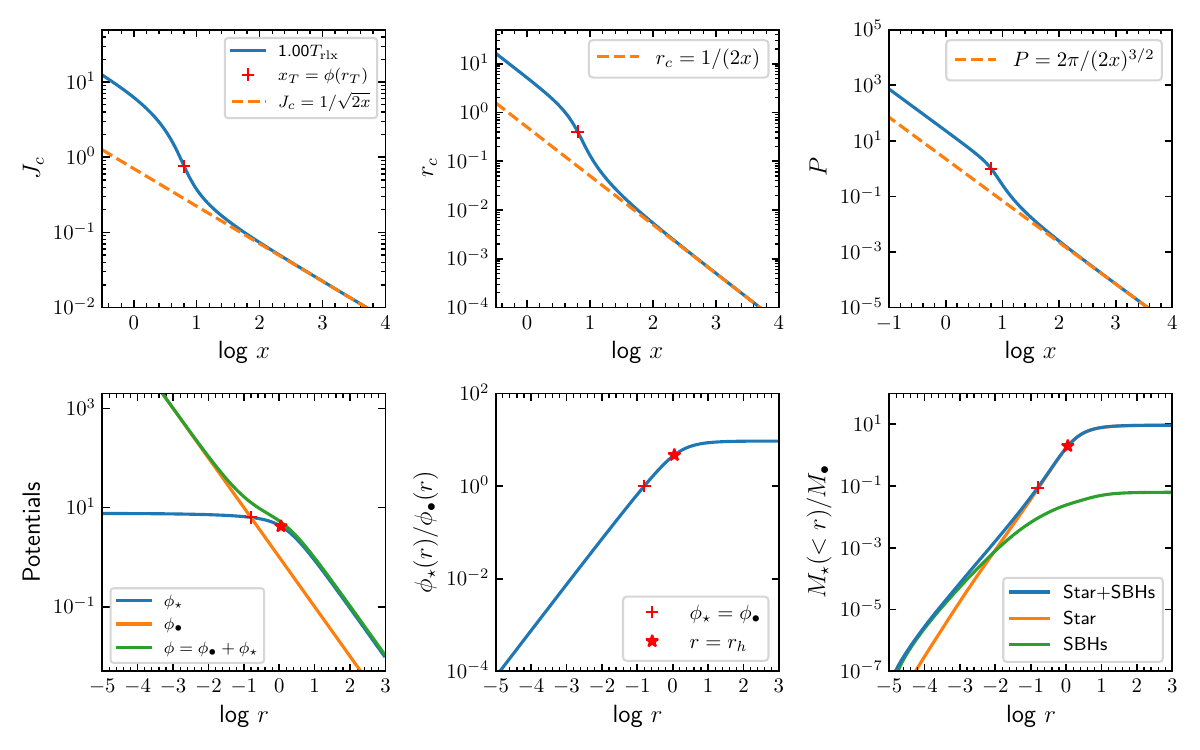}
	\caption{
Top panels: Angular momentum ($J_c$), radius ($r_c$), and orbital period ($P$) of circular orbits as a function of energy $x$. All quantities are expressed in the units described in Section~\ref{subsec:units}. The orange dashed line represents the expected values for Keplerian orbits. Bottom panels, from left to right: The potentials (stellar $\phi_\star$, MBH $\phi_\bullet$, and total $\phi$) as a function of radius; the ratio of $\phi_\star/\phi_\bullet$; and the enclosed stellar mass $M_\star$ relative to the mass of the MBH. The red cross in the bottom panels indicates the transition radius where $\phi_\star(r_T) = \bh/r_T$, with the corresponding energy $x_T = \phi_\star(r_T)$ shown in the upper panels. The red star in the bottom panels marks the position of the gravitational influence radius of the MBH ($M(<r_h) = 2\bh$). These results are taken from model M2.
	}
\label{fig:rcjc}
\end{figure*}

\begin{figure*}
	\center
	\includegraphics[scale=1.3]{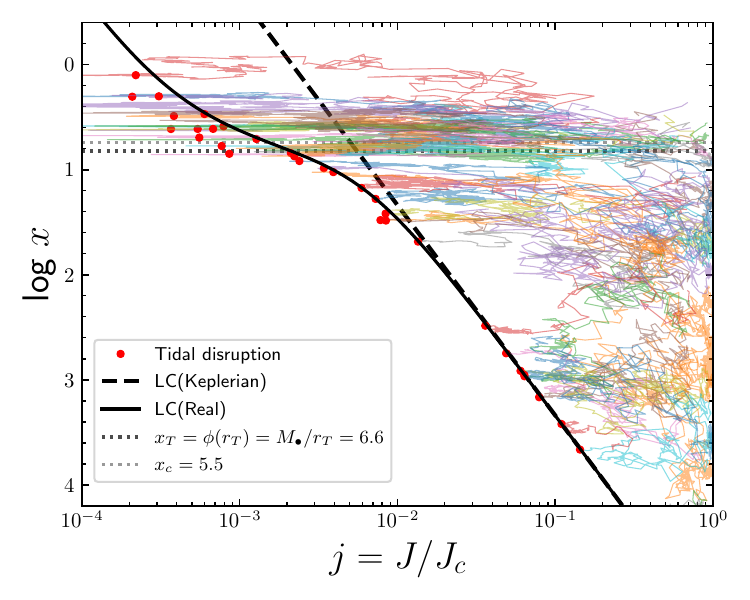}
	\caption{
 Examples of particle trajectories in the dimensionless energy ($x = E/\sigma_0^2$) and angular momentum ($j = J/J_c$) space, as calculated by \GNCc. The black solid (dashed) line represents the boundary of the loss cone when the stellar potential is included (ignored). A star is tidally disrupted if it moves within the loss cone and passes across the pericenter. The upper gray dotted line indicates the transition energy of the loss cone, $x_c$, as given by Equation~\ref{eq:rlc}, which separates the full and empty loss cone regions. The lower gray dotted line corresponds to the transition energy $x_T = \phi(r_T) = \bh/r_T$, where the stellar potential equals that of the MBH.
 }
	\label{fig:track}
\end{figure*}

In the previous version of the code, i.e., \GNCp~, although the stellar potential is ignored, the results should remain accurate in regions where the potential is dominated by the MBH. This is because these regions roughly correspond to the steady-state solution of the cluster, which has been widely studied and discussed in the literature~\citep[e.g.,][]{BW76,SM78,1978ApJ...226.1087C,2013ApJ...764...52B,ZA24}, and has also been extensively tested in~\citep{ZA24}. Therefore, it is important to verify whether \GNCc~reproduces the results of \GNCp~in the inner parts of the cluster when the mass growth of the MBH is ignored. Simultaneously, it is crucial to observe how the dynamics are modified in the presence of the stellar potential.

These results are discussed in Section~\ref{subsec:fixing_mbh_mass}. It is noteworthy that \GNCc~is capable of obtaining time-dependent dynamical solutions rather than merely approaching steady-state solutions, as the number of particles continues to decrease due to evaporation or the swallowing by the MBH.

For a more realistic study, the inclusion of MBH mass growth is essential, particularly in the early universe, where the majority of MBHs experience rapid mass growth alongside the dynamical evolution of the nuclear star cluster (NSC). In this initial study, we examine the simplest case in which the MBH grows its mass purely due to loss-cone accretion, such as the tidal disruption of stars and the direct swallowing of stars 
within event horizon or stellar-mass black holes within the last stable orbit. Further details are provided in Section~\ref{subsec:evolving_mbh_mass}.

\subsection{Non-evolving MBH mass}
\label{subsec:fixing_mbh_mass}

\begin{figure*}
	\center
	\includegraphics[scale=0.6]{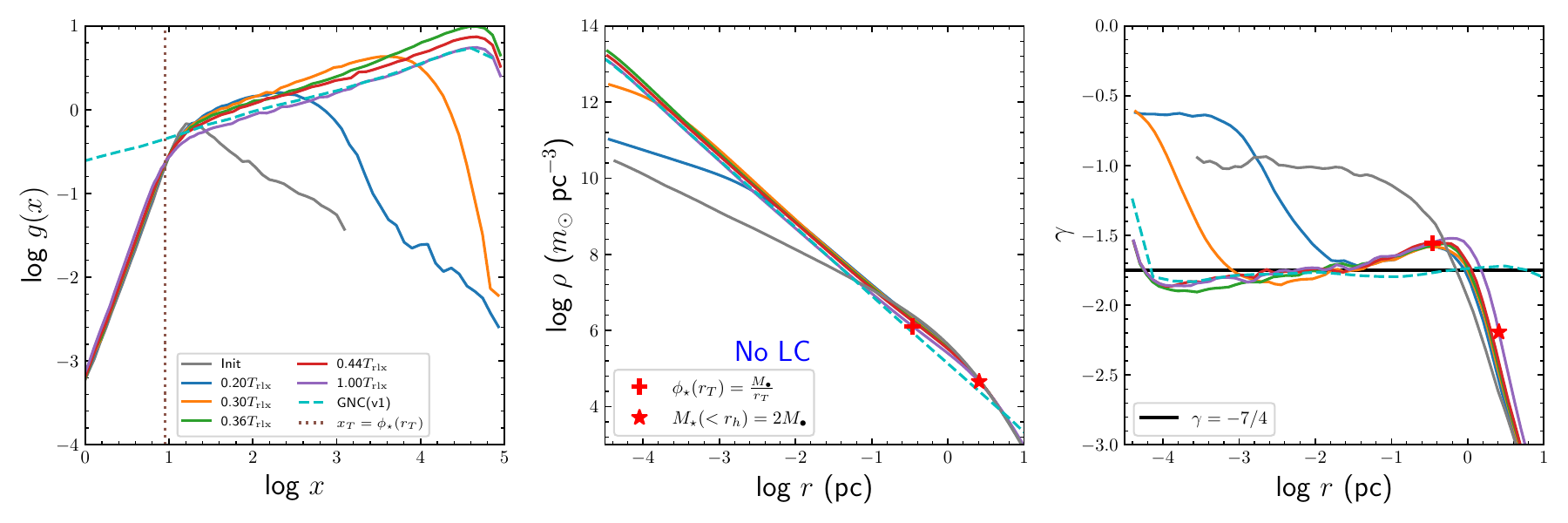}
	\includegraphics[scale=0.6]{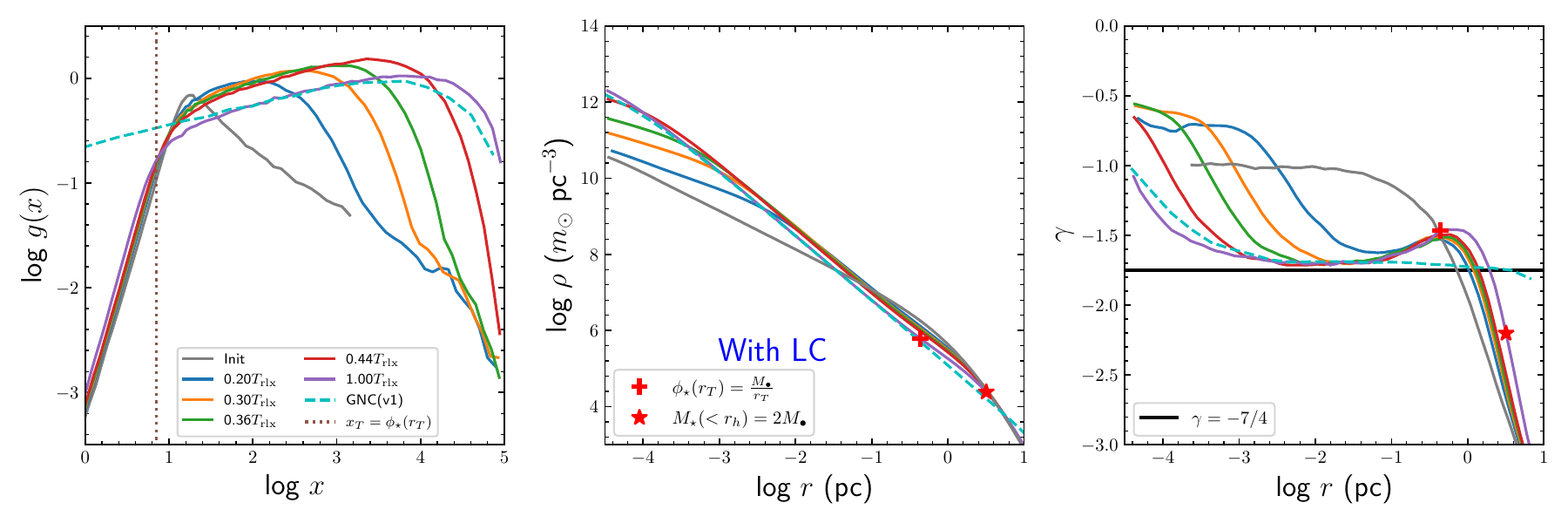}
	\caption{
 Dynamical evolution of a cluster (M1) consists of equal-mass stars and a central MBH obtained by \GNCc. The initial condition of M1 is listed in Table~\ref{tab:model}. Left panels: Dimensionless phase distribution $g(x)$, where $x=E/\sigma_0^2$. The dotted lines show the energy $x_T=\phi_\star(r_T)=\bh/r_T$, where $r_T$ is a critical transition radius where the stellar potential equals that of the MBH; Middle panels: Mass density $\rho$ as a function of radius $r$. The red cross shows the location of radius $r_T$ and the red star shows the position of $r_h$, i.e., where the enclosed mass equals twice the mass of MBH ($M_\star(<r_h)=2\bh$); Right panels: Slope index of the density profile $\gamma={\rm d}\ln \rho(r)/{\rm d}\ln r$. The top and bottom panels are results ignoring and including the effects of the loss cone, respectively. The cyan dashed lines in all panels are the steady-state results from \GNCp~which have ignored the effects of stellar potential. Note that the results of \GNCp~have been re-normalized such that its density equals the value of \GNCc~at $0.1r_T$ and at time $t=T_{\rm rlx}$.
	}
\label{fig:ev1_dehnen}
\end{figure*}
\begin{figure*}
	\center
	\includegraphics[scale=0.6]{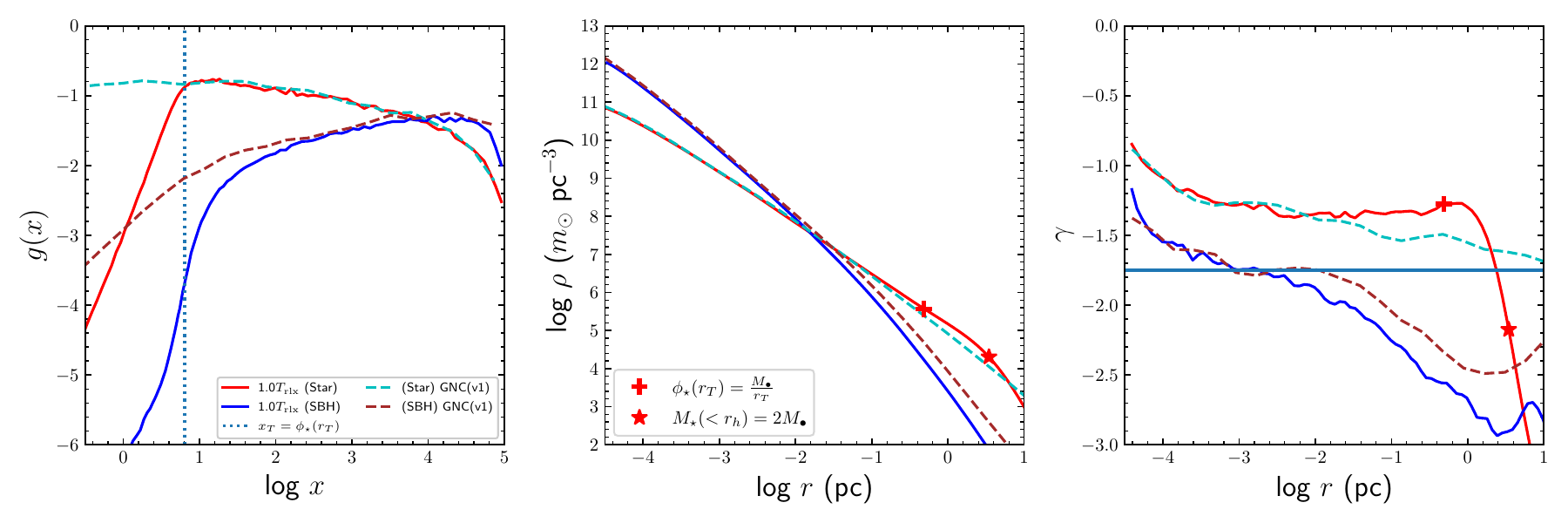}
	\caption{Similar to Figure~\ref{fig:ev1_dehnen} but for model M2, that consists of two 
	mass components (Stars+SBHs, see Table~\ref{tab:model} for its initial condition). Loss cone effects have been included.}
\label{fig:ev2_dehnen}
\end{figure*}

\begin{figure*}
	\center
	\includegraphics[scale=0.9]{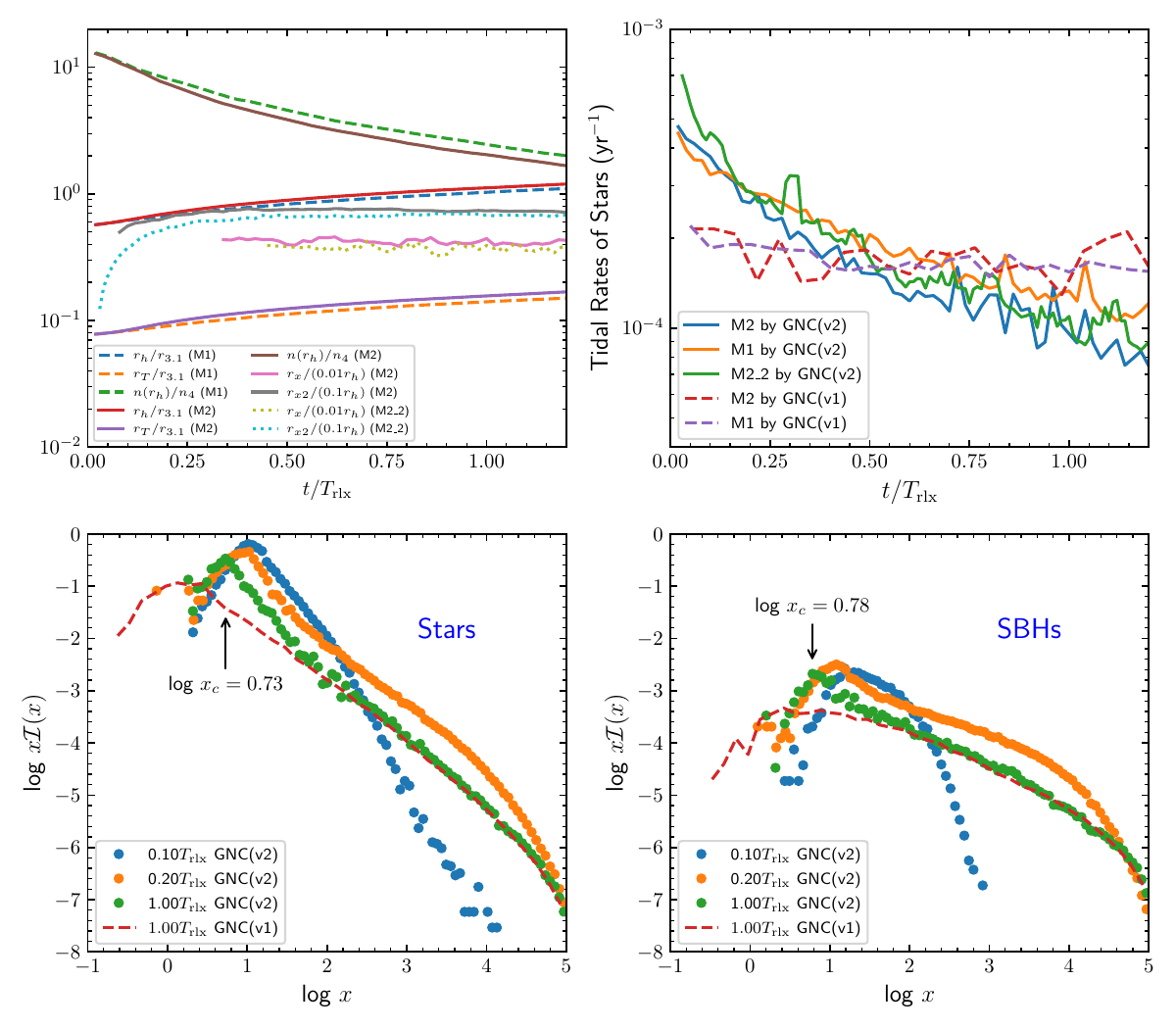}
	\caption{
 Top left panel: Evolution of characteristic radii of the cluster according to \GNCc: $r_h$ is the influence radius of the MBH; 
 $r_T$ is the transition radius where $\phi_\star(r_T)=\bh/r_T$; $r_x$ is the radius where the mass density of stars equals that 
 of the SBHs, i.e., $\rho_\star(r_x)=\rho_\bullet(r_x)$; $r_{x2}$ is the radius where $\rho_\star(r_x)m_\star=\rho_\bullet(r_x)m_\bullet$; 
 $r_{3.1}=3.1$ pc and $n_{4}=10^4$ pc$^{-3}$ are reference values for radius and number density, respectively. {The evolution of $r_T$, 
 $r_h$ and $n_4$ in model M2\_2 is very similar to those of model M2, and thus are not shown here for clarity.}
 Top right panel: Evolution of the tidal disruption rates of stars in models M1 and M2, as calculated by \GNCc~or \GNCp. Bottom left and right panels: Dimensionless flux of stars (left panel) and SBHs (right panel) falling into the loss cone in model M2. $x_c$ is the energy separating the full and empty loss cone regions according to Equation~\ref{eq:qx}. The red dashed lines in the bottom panels represent the results from model M2 in Table 1 of~\citet{ZA24} as calculated by \GNCp.
	}
\label{fig:td_rate_fix}
\end{figure*}

In this section, we present the evolution and density profile of the nuclear star cluster (NSC) when the mass of the central MBH is held constant. We also compare these results with those obtained from the previous version, \GNCp. Since the stellar potential is ignored in \GNCp, this comparison allows us to illustrate how the evolution and dynamics are altered in the presence of the stellar potential. 

{Similar to \GNCp, in the simulations conducted in this section, the interval is set to a small fraction ($0.005\sim0.01$) of the two-body relaxation timescale, \(T_{\rm rlx}\), estimated at the influence radius of the MBH at the start of the simulation. Although the two-body relaxation time varies across the cluster, it is typically a weak function of distance in the presence of a central MBH (e.g., see Figure 1 of~\citet{Zhang19}). Consequently, these intervals are sufficiently short to ensure the convergence of the simulation results.}

We discuss {three models (M1, M2 and M2\_2, as outlined in Table~\ref{tab:model})}, one consisting of stars with equal mass and the other two additionally containing a fraction of $10^{-3}$ stellar-mass black holes (SBHs) with mass $m_\bullet=10\,\msun$. The initial density profile for M1 and M2 is based on the Dehnen model with $\gamma_i=1$, $M_{\rm cl}(0)/\bh=10$, $r_a=0.7r_0$, and $m_0=\bh=4\times 10^6\,\msun$, where $r_0=3.1$ pc. {M2\_2 is similar to M2 but with $\gamma_i=1.5$; the value of $r_a$ is set to $0.94r_0$ such that it has the same initial effective radius of cluster 
as those of M2.}

In \GNCc, we include the effects of the loss cone, defined as the tidal radius for stars ($R_t$) or when the pericenter of the orbit $r_p$ equals $8r_g=\frac{8G\bh}{c^2}$ for SBHs. The tidal radius of a star can be determined by

\be
R_t=\eta\left(\frac{\bh}{m_\star} \right)^{1/3}R_\star,
\ee

\noindent
where $\eta$ is a parameter that varies depending on the structure of the star or relativistic effects. Previous studies have adopted values of $\eta \sim 0.2-4$~\citep{1999MNRAS.309..447M, 2012PhRvD..85b4037K, 2020ApJ...904...98R}. 
{If the tidal radius or the orbital pericenter is within the event horizon, i.e., $R_t<2r_g$ or $r_p<2r_g$, 
it produces a non-flaring event, rather than a tidal disruption. The star is swallowed whole.}
Here, we adopt $\eta=1$ for simplicity, although it is important to note that the rates of tidal disruption events can be highly sensitive to the value of $\eta$ in the mass range $\bh=10^7\sim 10^8\,\msun$. 

We run simulations using \GNCp~adopting models M1 and M2 from Table 1 of~\citet{ZA24} to compare with the corresponding results of the models M1 and M2 presented here. In the previous version, \GNCp, the outer parts of the cluster effectively assume an isothermal distribution, which differs significantly from the simulations conducted here. Therefore, the comparisons between the results of these two versions are only meaningful in the inner regions where the MBH dominates the dynamics.

The dynamics transition from being dominated by the MBH to being dominated by the stellar components around a characteristic radius $r_T$ or a characteristic energy $x_T$, given by $x_T=\phi_\star(r_T)=\bh/r_T$. $r_T$ (or $x_T$) corresponds to the radius (or energy) where the potential of the MBH equals that of the stellar objects, which can be determined using \GNCc. The results of \GNCp~are then re-normalized so that the density $\rho$ at $0.1r_T$ from \GNCp~matches that of \GNCc~at $1~T_{\rm rlx}$, where $T_{\rm rlx}$ is the relaxation time defined at the influence radius of the MBH at the beginning of the simulation.

From Figure~\ref{fig:rcjc}, it is evident that stellar orbits change significantly around the transition energy $x_T$. Below this energy, the orbits follow Keplerian dynamics. Above $x_T$, the radius $r_c$, angular momentum $J_c$, and period $P$ of the circular orbits are about one or two orders of magnitude larger than those in the Keplerian case, simply because the enclosed mass of stellar objects is much larger than the mass of the MBH.

\subsubsection{NSC evolution}

Figure~\ref{fig:track} illustrates examples of particle trajectories in model M1 as simulated by \GNCc. Particles move within the energy and angular momentum space through stochastic processes governed by diffusion coefficients. Notably, in the outer regions of the cluster, the size of the loss cone is modified in the presence of the stellar potential, as clearly shown by the thick black solid line in Figure~\ref{fig:track}. This modification occurs because, for a given energy, the angular momentum for a circular orbit is determined by the enclosed mass of the cluster within that orbit, which is significantly larger than the mass of the MBH (see also Figure~\ref{fig:rcjc}). As a result, the dimensionless angular momentum $j=J/J_c$ is much smaller than in the Keplerian case (represented by the black dashed line). This difference in the size of the loss cone occurs in regions where $x\lesssim x_T\sim 6.6$. 

In the full loss cone region, particles can frequently move in and out of the loss cone before being destroyed by the MBH (i.e., before passing through the pericenter). The full loss cone region is located at $x\lesssim x_c\sim 5.5$, where $x_c$ will be further explained in Section~\ref{subsec:ls_rate}. As shown in Figure~\ref{fig:track}, the full loss cone region, where $x\lesssim x_c\sim x_T$, is dominated by the enclosed stellar components rather than the MBH. These results suggest that it is crucial to include the stellar potential to accurately study the dynamics in the full loss cone region.

Figure~\ref{fig:ev1_dehnen} shows the evolution of the phase distribution function ($g(x)$), mass density ($\rho$), and the slope index of the density profile ($\gamma$) for model M1, which consists of stars of equal mass ($m_\star=1\msun$). The transition in the phase distribution function occurs near the energy $x_T=\phi_\star(r_T)$. As here density profile does not follow isothermal distribution, $r_T$ is approximately 10 times smaller than the influence radius of the MBH (defined by $M_\star(<r_h)=2\bh$), which is marked by the red star in the middle panel. From Figure~\ref{fig:ev1_dehnen}, it is evident that the cluster is in a quasi-steady state at $\simeq 0.4\sim 1 T_{\rm rlx}$, after which the density of the entire cluster continuously and slowly decreases, primarily due to the depletion of particles, such as through tidal disruption of stars.

Comparing \GNCc~with \GNCp~for both models M1 and M2 in Figures~\ref{fig:ev1_dehnen} and~\ref{fig:ev2_dehnen}, it is clear that \GNCc~successfully reproduces the results of \GNCp~for the density profiles within $\sim 0.1r_T$, regardless of whether the effects of the loss cone are included.

We also investigate how some characteristic quantities closely related to the cluster dynamics evolve over time. In addition to $r_T$ and $r_h$, we are interested in the stellar number density $n(r_h)$ at $r_h$, the radius where the mass density of stars equals that of the SBHs ($r_x$, defined by $\rho_\star(r_x)=\rho_\bullet(r_x)$), and the radius where the contribution to the two-body relaxation time from stars equals that from the SBHs~\citep[$r_{x2}$, defined as $\rho_\star(r_{x2})m_\star=\rho_\bullet(r_{x2})m_\bullet$ in][]{2019PhRvD..99l3025A}.

The top left panel of Figure~\ref{fig:td_rate_fix} shows the evolution of these quantities as a function of time in models M1 and M2. 
It is noteworthy that both $r_h$ and $r_T$ gradually increase to approximately 1.5 times their initial values after $1T_{\rm rlx}$. The density at $r_h$, i.e., $n(r_h)$, decreases to about 20\% of its initial value. This indicates that the cluster continues to expand due to dynamical relaxation. The cross radius $r_x$ is approximately $4\times 10^{-3}r_h$, and the SBHs dominate the relaxation inside a radius of $r_{x2}\sim 0.07 r_h$. The ratios of these two radii with respect to $r_h$ evolve only slightly after $t\gtrsim 0.5T_{\rm rlx}$.
{The results of M2\_2 show similar evolution to those of M2, although $r_{x2}$ is much smaller at the beginning.} 

\subsubsection{Loss cone rate}
\label{subsec:ls_rate}

As the boundary conditions in \GNCp~and \GNCc~are different (see Section~\ref{subsec:bd_condition}), a direct comparison of the loss cone infall rates between these two versions is not appropriate. Similar to the approach taken in the previous section, we first re-normalize the weighting of each particle in \GNCp~by matching the density of stellar components to those in \GNCc~at a radius of $0.1r_T$. After this correction, the loss cone rates from \GNCp~can be meaningfully compared with those from \GNCc.

The evolution of the loss cone (tidal disruption) rate of stars is shown in the top right panel of Figure~\ref{fig:td_rate_fix}. We observe that the tidal disruption rates from \GNCp~decrease slightly at the beginning of the simulation, after which they quickly reach a steady state. 
Including the stellar potential and adopting the evaporating boundary condition 
in \GNCc~results in a continuously decreasing rate of tidal disruption. 
In model M2, due to mass segregation, the SBHs sink into the inner regions while the stars tend to move 
outward, leading to slightly lower tidal disruption rates compared to the scenario where the cluster consists only of stars.
{In model M2\_2, the initial density in the inner regions is higher, resulting in rates that are approximately $50\%$ greater than those of model M2. However, the rates in M2\_2 quickly converge to those of M2 after approximately $0.1T_{\rm rlx}$.}

It is also interesting to examine how the flux of objects into the loss cone evolves in the full or empty loss cone regions. In \GNCp, it was demonstrated in Appendix C of~\citet{ZA24} that the rate of stellar objects falling into the loss cone is given by

\be\ba
R_{\rm lc}&=\frac{4\sqrt{2}\pi^2}{3}r_h^{9/2}n_h^{2}(\bh G)^{-3/2}\ln \Lambda (Gm_\star)^2 I_0\\
=&3.86\times 10^{-5}{~\rm yr}^{-1} \times  
r_{3.1}^{\frac{9}{2}}n_{4}^{2}
M_{4e6}^{-\frac{3}{2}} \Gamma_{15.2}
m_1^2 I_0
\label{eq:rlc}
\ea\ee
where $r_{3.1}=r_h/(3.1~{\rm pc})$, $n_{4}=n_h/(10^4~{\rm pc}^{-3})$,
$M_{4e6}=\bh/(4\times 10^6 M_\odot)$, $m_1=m_\star/(1~M_\odot)$ and $\Gamma_{15.2}=\ln \Lambda/15.2$.
$I_0$ is a dimensionless rate that is given by
\be
I_0=\int^{x_{\rm max}}_0 \mathcal{I}(x)dx,
\ee
where $\mathcal{I}(x)$ is a dimensionless flux of particles destroyed by loss cone at energy $x$, 
$x_{\rm max}$ is the energy at the higher boundary (See Section~\ref{subsec:bd_condition}).

When including the stellar potential, the critical energy $x_c$ separating the full 
and empty loss cone is given by~\citep{1977ApJ...211..244L,ZA24}
\be
q(x)=\frac{D_{JJ,0}P(x)}{j_{lc}^2}\simeq -0.36\ln j_{\rm lc},
\label{eq:qx}
\ee

\noindent
where $D_{JJ,0}=D_{JJ}(J\rightarrow 0)$ is the diffusion coefficient of angular momentum near the edge of the loss cone, $P(x)$ is the orbital period at energy $x$, $j_{\rm lc}=J_{\rm lc}/J_c(x)$, and $J_{\rm lc}=\sqrt{2\bh r_p}$ is the size of the loss cone, with $r_p$ being the pericenter of the orbit at energy $x$ (equal to the tidal radius of stars or $8r_g$ for SBHs). By solving the above equation, the critical energy $x_c$ can be obtained.

The bottom panels of Figure~\ref{fig:td_rate_fix} display the dimensionless flux in model M2 for both stars and SBHs. Note that when using \GNCc~to obtain $\mathcal{I}(x)$ from Equation~\ref{eq:rlc}, it is necessary to adopt the instantaneous values of $r_h$ and $n_h$ at the given time of the simulation, as these values vary over time. \GNCc~and \GNCp~show very good consistency in the empty loss cone region ($x\gg x_c$) after approximately $1T_{\rm rlx}$ for both stars and SBHs.

The position of $x_c$, determined by solving Equation~\ref{eq:qx}, is indicated by black arrows in the bottom panels of 
Figure~\ref{fig:td_rate_fix} for both stars and SBHs. The turning point of the flux $\mathcal{I}(x)$ from the \GNCc~simulation aligns 
well with the calculated value of $x_c$. The results for $\mathcal{I}(x)$ from \GNCc~and \GNCp~diverge around $x_c$, which is expected, 
as the inclusion of the stellar potential in \GNCc~alters the dynamics in the outer parts of the cluster. Notably, the transition energy $x_c$ 
is approximately $0.5$ dex lower than when the stellar potential is ignored.

\subsection{If evolving the mass of MBH}
\label{subsec:evolving_mbh_mass}
\begin{figure*}
	\center
	\includegraphics[scale=0.7]{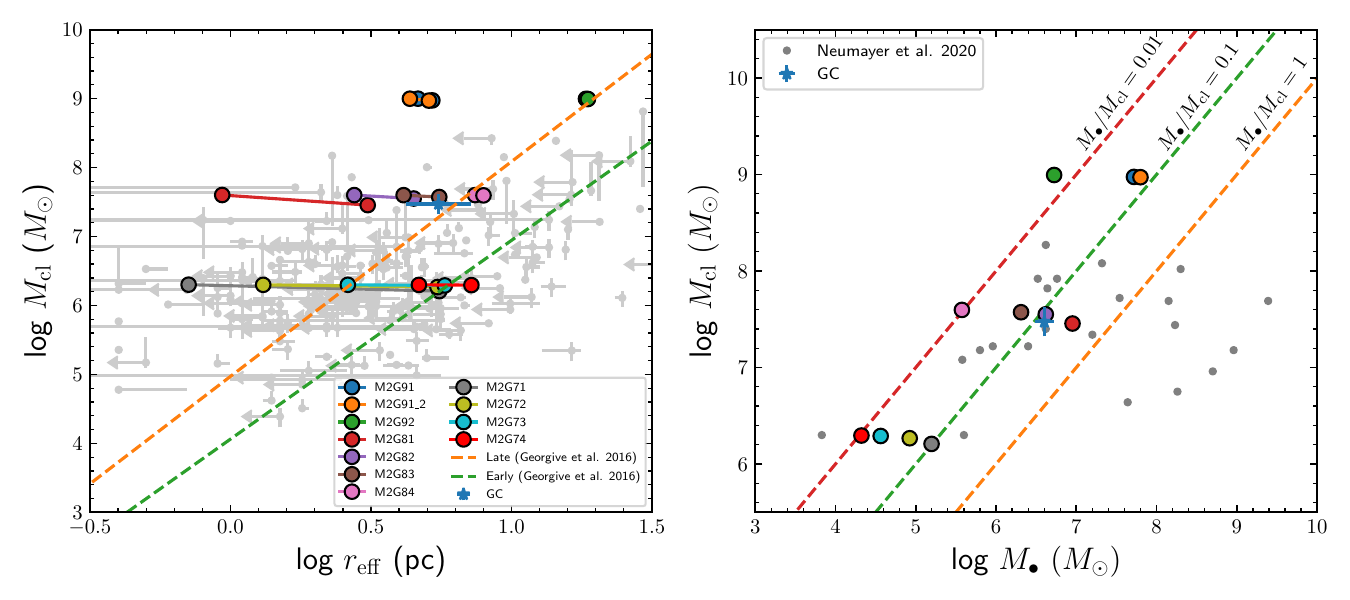}
	\caption{
 Left panel: Cluster mass $M_{\rm cl}$ and effective radius $r_{\rm eff}$ for the M2G series of models in Table~\ref{tab:model}. 
 For each model, the point on the left represents the initial value, and the point on the right represents the value at 12 Gyr. 
 The gray dots with error bars indicate observations of nearby galaxies from~\citet{2016MNRAS.457.2122G}. 
 The dashed line represents the fitted correlation between $M_{\rm cl}$ and $r_{\rm eff}$ by \citet{2016MNRAS.457.2122G} 
 for both late and early-type galaxies. Right panel: Final cluster mass $M_{\rm cl}$ and the mass of the MBH $\bh$ for 
 each model. Gray dots represent data from~\citet{2020A&ARv..28....4N}. The mass of the Milky Way MBH is taken 
 from~\citet{2009ApJ...692.1075G}. Dashed lines indicate reference lines where $\bh/M_{\rm cl}=0.01$, $0.1$, and $1$. 
 {The results of M2G82\_2 and M2G82m are very similar 
 to those of M2G82, M2G71m. On the other hand, M2G71\_2 is very similar to M2G71}, thus are not shown here for clarity.
}
\label{fig:mbh_varying_ini}
\end{figure*}

\begin{figure*}
	\center
	\includegraphics[scale=0.9]{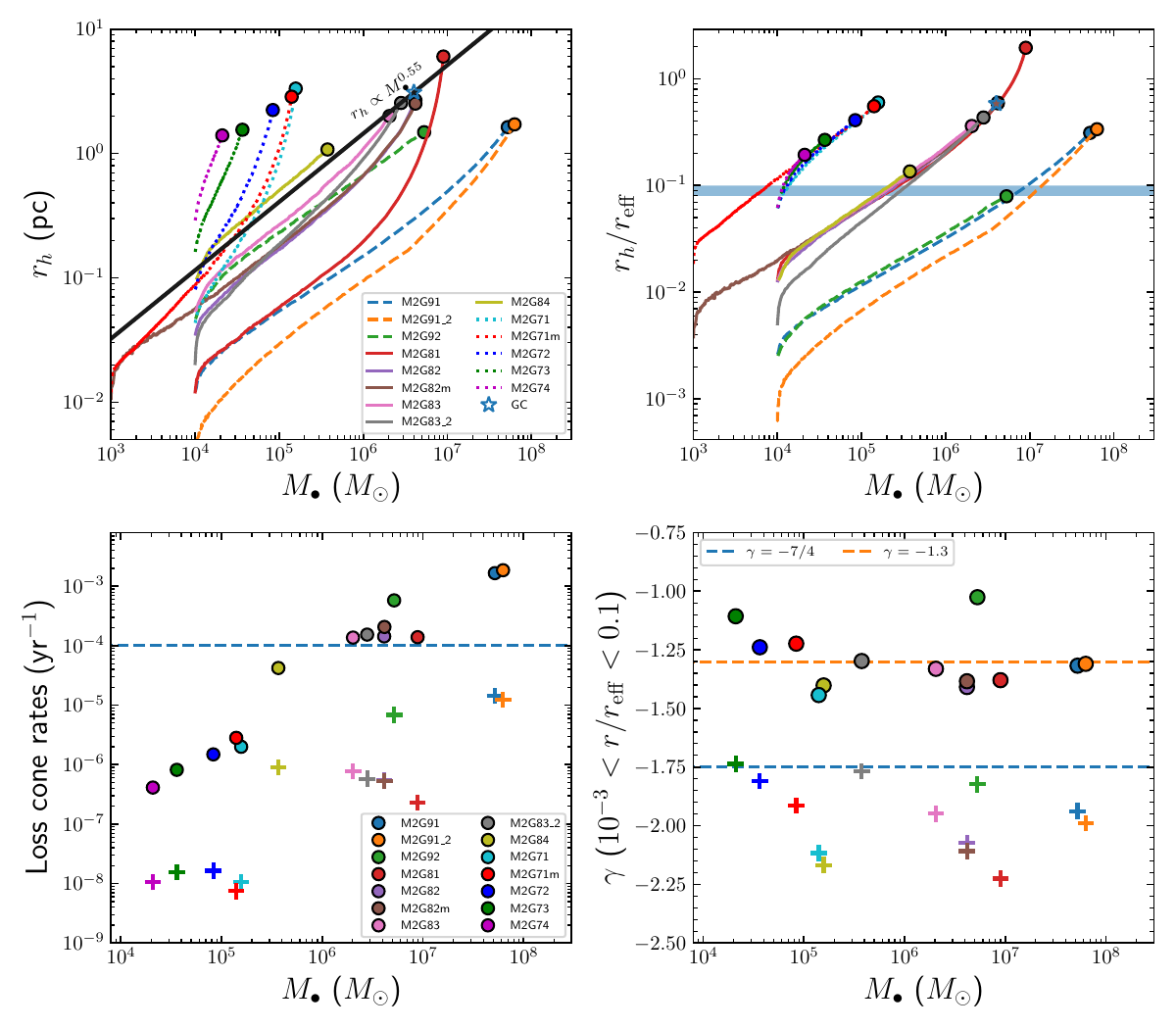}
	\caption{
 Top Left panel: Evolution of the influence radius of the MBH ($r_h$, defined by $M(<r_h)=2\bh$) 
 and the mass of the MBH ($\bh$) over 12 Gyr for different models. The black line represents the expectations from the $\bh-\sigma$ relation, i.e., $r_h=\bh/\sigma^2$, where the $\bh-\sigma$ relation is provided by~\citet{2013ARA&A..51..511K}. Top right panel: Evolution of $r_h/r_{\rm eff}$ and the mass of the MBH ($\bh$).
 The shaded light-blue region marks where $r_h/r_{\rm eff}=0.08\sim 0.1$.
 The lines and symbols are the same as the top left panel.
 Bottom left panel: Loss-cone accretion rates versus the present-day mass of the MBH, 
 with colored filled circles indicating tidal disruption rates of stars and crosses of the same color showing the swallowing rates of SBHs. The vertical dashed line indicates a reference rate of $10^{-4}$ yr$^{-1}$. Bottom right panel: Slope index of the density of stars (filled circles) and SBHs (crosses of the same color) versus the present-day mass of the MBH. The slope indices are calculated based on the densities between $10^{-3}r_{\rm eff}<r<0.1r_{\rm eff}$. The horizontal dashed lines represent reference indices of $-1.3$ and $-1.75$.
 {In all panels, the results of model M2G71\_2 are very similar to those of M2G71, and hence not shown for clarity.}
}
\label{fig:mbh_varying_ini_rh}
\end{figure*}

\begin{figure*}
	\center
	\includegraphics[scale=0.6]{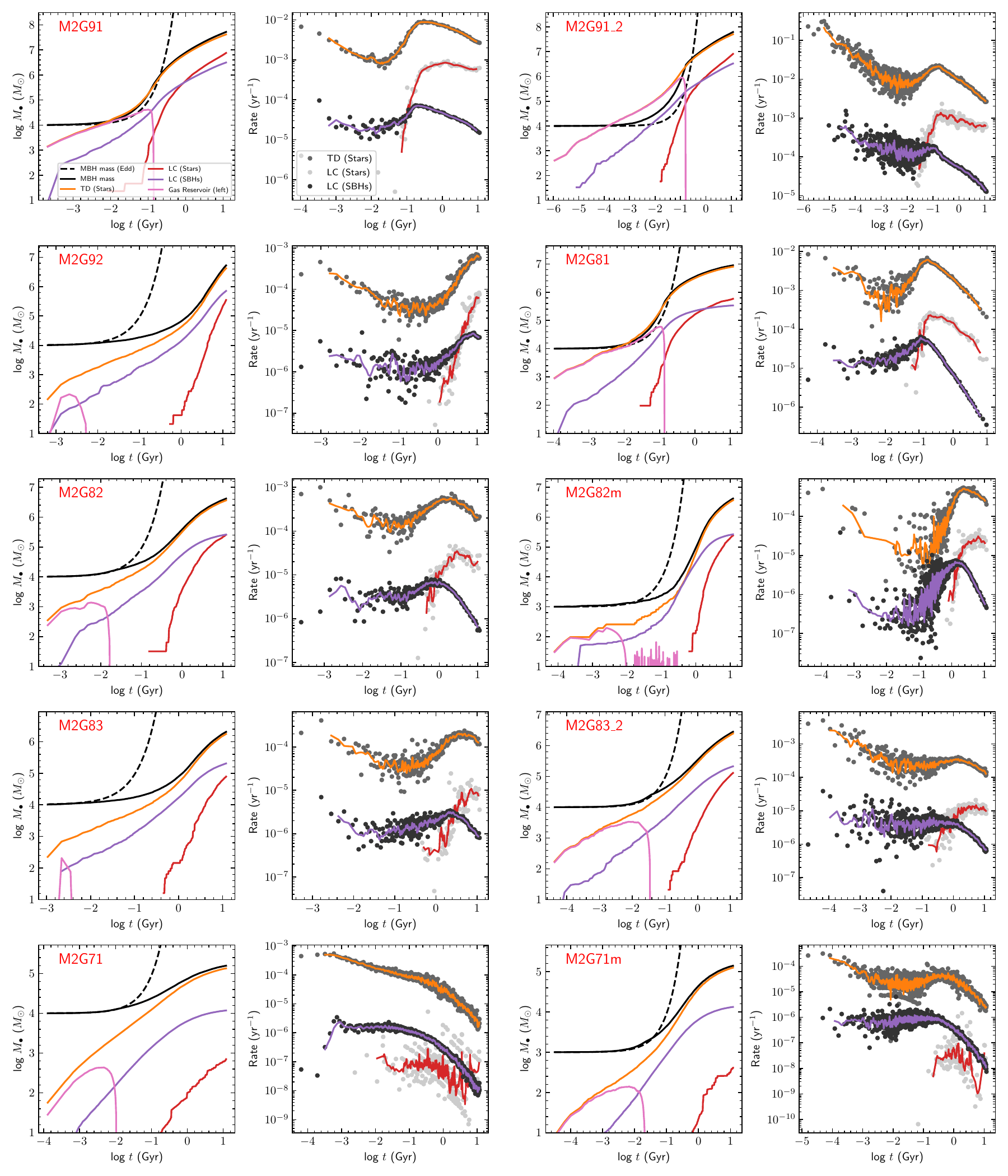}
	\caption{
 Evolution of the MBH mass and the rates of loss-cone accretion for some of 
	M2G model series in Table~\ref{tab:model}. The first and the third columns 
	show the mass growth (black solid line), {theoretical growth of mass under Eddington limit (black dashed line),
	the current mass of gas reservoir (``Gas reservoir (left)'')}, the cumulative contribution from tidal disruption of stars 
	(``TD (stars)''), stars falling into the horizon of MBH without producing observable flares 
	(``LC (stars)''), and loss-cone accretion of SBHs (``LC (SBHs)''). The second and the fourth columns show the corresponding evolution of the loss-cone accretion rates. The dots correspond to the 
 rates in each iteration of the simulation, and lines with the same color of those in the 
 first panel show the rates (smoothed via moving average for every five data points).
}
\label{fig:mbh_varying_evl}
\end{figure*}

In this section, we proceed to investigate the evolution of nuclear star clusters (NSCs) when the mass of the MBH is growing. For simplicity, we consider only the accretion of material that falls into the loss cone of the MBH. While the evolution of NSCs can become increasingly complex by incorporating additional factors such as stellar collisions and gaseous material accretion into the MBH, our primary focus in this first application of \GNC~is to examine the effects of the loss cone. Therefore, we defer these additional complexities to future studies.

In \GNCc, when a star passes through the pericenter within its tidal disruption radius, the star is destroyed, and its entire mass is added to a gas reservoir that can be accreted by the central MBH at a rate given by
\be
\dot M_{\rm edd}=2.22\msun\,{\rm yr}^{-1}~\times \frac{0.1}{\epsilon}\frac{\bh}{10^8\msun}
\ee
where we adopt $\epsilon=0.1$ as the radiative efficiency.

In reality, about half of the star's mass is directly accreted by the MBH, while the remaining mass is expelled and becomes unbound to the MBH. However, it is important to note that some of this expelled gas may remain bound to the NSC or the bulge of the galaxy, with a portion of it potentially falling back and eventually being accreted by the MBH.

{If the pericenter of a star  or the tidal radius of a star 
is within the event horizon of the MBH, i.e., $r_p<2r_g$ or $R_t<2r_g$ (assuming that it is no spinning), or if a SBH  
is within the last stable orbit, i.e., $r_p<8r_g$, the mass of star or SBH is then instantaneously 
added into the mass of MBH when they pass through the pericenter.}

We perform a series of simulations using \GNCc~with a seed MBH of $\bh=10^3\,\msun$ or $10^4\,\msun$. These models belong to the M2G series, details of which can be found in Table~\ref{tab:model}. Each model is run for a Hubble time, i.e., 12 Gyr. These models are designed so that, by the end of the simulation, the total mass of the stellar cluster $M_{\rm cl}$ and the effective radius $r_{\rm eff}$ roughly match the recent observations of NSCs in the Milky Way or nearby galaxies~\citep[][See Figure~\ref{fig:mbh_varying_ini} and~\ref{fig:dehnen_gc}]{2003ApJ...594..812G,2016MNRAS.457.2122G,2018A&A...609A..27S,2019MNRAS.484.1166M,2020A&ARv..28....4N}. The effective radius is defined by $S(r_{\rm eff})=M_{\rm cl}/2$, where $S(R)=2\pi \int^R_0 \Sigma(R)RdR$ is the cumulative surface density, and $\Sigma$ is the surface density of the cluster, defined by
\be
\Sigma(r)=2\int^\infty_r \frac{\rho(R)RdR}{\sqrt{R^2-r^2}}
\ee

{ In this section, the simulation intervals are determined based on both the two-body relaxation timescale and the accretion timescale. To ensure adiabatic adaptation of particle energy, as described in Section~\ref{subsec:including_ait}, the potential of the MBH must vary slowly, requiring intervals small enough to prevent significant growth of the MBH mass. }

{Given the $e$-folding timescale for MBH mass growth due to accretion, \(T_{\rm acc}=\bh/\dot M_{\rm edd}\simeq 45\) Myr, and a restriction that the mass growth of \(\bh\) should not exceed a fraction \(f_{\rm acc}\) of its current mass, the interval is defined as:}

\be
\delta t=
\left\{
	\begin{array}{cc}
	{\rm min}(f_{\rm acc} T_{\rm acc},f_{\rm rlx}T_{\rm rlx}),& {\rm if~} M_{\rm gas}>f_{\rm acc}\bh\\
	f_{\rm rlx}T_{\rm rlx},& {\rm if~} M_{\rm gas}<f_{\rm acc}\bh
	\end{array}
	\right.
\ee

{Here, \(M_{\rm gas}\) is the current mass of the gas reservoir remaining after the previous iteration. We set \(f_{\rm acc}=0.1\), ensuring that within each interval, the MBH mass does not grow by more than \(10\%\) of its current mass. }

{Similarly to Section~\ref{subsec:fixing_mbh_mass}, \(T_{\rm rlx}\) is the two-body relaxation timescale estimated at the MBH influence radius, with \(f_{\rm rlx}=0.005-0.01\). Note that \(T_{\rm rlx}\) is updated after each simulation iteration to reflect the updated MBH mass.
}

Detail discussions of the model results are as follows.

\subsubsection{Size expansion of NSCs}
\label{subsec:size_expansion_of_NSCs}
We can compare the effective size at $12$Gyr for different models with those of
the observed late-type galaxies~\citep{2016MNRAS.457.2122G},
\be
\log \left(\frac{r_{\rm eff}}{3.31~\rm pc}\right)=0.321 \log \left(\frac{M_{\rm cl}}{3.6\times 10^6\msun}\right)-0.011,
\ee
and early-type galaxies~\citep{2016MNRAS.457.2122G}: 
\be
\log \left(\frac{r_{\rm eff}}{6.27~\rm pc}\right)=0.347 \log \left(\frac{M_{\rm cl}}{1.95\times 10^6\msun}\right)-0.024.
\ee
The results are shown in the left panel of Figure~\ref{fig:mbh_varying_ini}.

The effective radius of a NSC shows noticeable expansion if its mass is approximately equal to or smaller than that of the Milky Way NSC ($M_{\rm cl} \leq 4\times 10^7\,\msun$) and it initially has a size of $\leq 10$ pc. For Milky Way-like NSCs (models M2G8 series), their sizes expand by approximately 1.5 to 3 times their initial size, which falls within the range suggested by observations of late-type galaxies. For smaller NSCs ($M_{\rm cl}=10^6\,\msun$, models M2G7 series), the size expansion over 12 Gyr is so significant that their current sizes are more consistent with those of NSCs in early-type galaxies. Massive NSCs (e.g., $M_{\rm cl}>10^9\,\msun$) have relatively longer relaxation times, which typically result in minimal size expansion over cosmic time. {These results are consistent with the size expansion of NSCs explored in~\citet{2009ApJ...694..959M}.}

These findings suggest that many NSCs, which are similar in size or smaller than the Milky Way NSC, may have been much more compact in the early universe and have undergone significant size expansion over cosmic time. For the currently observed small NSCs in late-type galaxies (i.e., $M_{\rm cl}\sim 10^6\,\msun$), some of them may have formed more recently, as their sizes would likely be much larger than the observed values if they had formed in the early universe.
  
\subsubsection{Cosmological evolution of tidal disruption rates}

We find that the higher the initial density in the inner regions of the cluster, 
the higher the rate of loss-cone accretion, and the more substantial the growth 
of the MBH over 12 Gyr. Generally, the tidal disruption rates of stars at 12 Gyr 
are approximately $10^{-4}\sim 10^{-3}$ yr$^{-1}$ for massive NSCs ($M_{\rm cl}=10^9\,\msun$), 
$10^{-5}\sim 10^{-4}$ yr$^{-1}$ for Milky Way-like NSCs ($M_{\rm cl}=4\times 10^7\,\msun$), 
and $10^{-7}\sim 10^{-6}$ yr$^{-1}$ for smaller NSCs ($M_{\rm cl}=2\times 10^6\,\msun$). 
Detailed rates for each model can be found in the last column of Table~\ref{tab:model} or the 
bottom left panel of Figure~\ref{fig:mbh_varying_ini_rh}.

From Figure~\ref{fig:mbh_varying_evl}, it appears that the evolution of the loss-cone accretion depends on the initial conditions of the cluster and the growth history of the central MBH. For each model, the rates can vary significantly over 12 Gyr.

We find that the loss-cone accretion rate  as a function of time depends on the evolution of the ratio $r_h/r_{\rm eff}$ 
(see top right panel of Figure~\ref{fig:mbh_varying_ini_rh} for the evolution as function of MBH mass, 
where $r_h$ is defined by $M(<r_h)=2\bh$). If initially $r_h\ll 0.08\sim 0.1r_{\rm eff}$ (such as in the M2G9 and M2G8 series), i.e., when the gravitational influence radius of the MBH is much smaller than the size of the cluster, the evolution of the rates occurs in three distinct phases:

\begin{enumerate}
    \item In the early universe, loss-cone accretion begins at a high rate, which gradually decreases. This occurs because most stellar objects in the NSC are initially weakly bound or unbound to the MBH, with their orbits dominated by stellar potentials. The empty loss cone region is almost entirely devoid of stars, while most stars are located in the full loss cone region. Consequently, at the start of the simulation, loss-cone accretion events primarily involve the consumption of stars near or inside the full loss cone region of the MBH.
    \item After a brief period, the rates begin to increase. As more stars start sinking into the central regions due to adiabatic invariant theory—i.e., the response of the stars' orbits to the increasing potential of the MBH (see Section~\ref{subsec:including_ait})—the empty loss cone region begins to fill with stars. As the mass of the MBH grows, the size of the loss cone also increases, leading to progressively higher rates of loss cone events.
    \item The rates reach a peak when $r_h$ increases to approximately $0.08r_{\rm eff}$ and then start to decrease when $r_h \gtrsim 0.1-0.2r_{\rm eff}$. This decline occurs because the gravitational influence radius of the MBH becomes comparable to the size of the cluster, leading to the gradual consumption of stars in both the full and empty loss cone regions. As a result, the rates will gradually decrease.
\end{enumerate}

On the other hand, if initially $r_h \gtrsim 0.08r_{\rm eff}$ (as in the M2G7 series), the rates consistently decrease over time. This occurs because, in this scenario, the MBH is initially massive enough that both the empty and full loss cone regions are already filled with stars, causing the system to evolve directly into phase (3) described above.

The tidal disruption of stars contributes to the majority of the MBH's mass increment over 12 Gyr, typically being one or two orders of magnitude larger than the mass contribution from SBHs or stars directly swallowed by the MBH.

{As shown in Figure~\ref{fig:mbh_varying_evl}, model M2G91\_2 exhibits significantly higher tidal disruption rates in the early Universe ($\lesssim 100$ Myr) compared to M2G91. This is primarily due to the higher initial density slope (\(\gamma_i=1.5\)) in M2G91\_2, as opposed to \(\gamma_i=1\) in M2G91, in the inner region of the cluster. Both models share the same effective cluster size, so the observed differences arise from the variation in slope rather than normalization.}

{Similar trends are observed when comparing models M2G82\_2 and M2G82, or M2G71\_2 and M2G71, though the differences in those cases are smaller.
}
\subsubsection{Comparing the simulated present-day tidal disruption rates with observations}
{The tidal disruption rates in steady-state solutions (without evolving the MBH's mass) due to relaxation typically yield a rate of \(\sim 10^{-4}\,{\rm yr}^{-1}\), which decreases with the MBH mass as \(\dot{R} \propto \bh^\alpha\), where \(\alpha \simeq -0.3\)~\citep[e.g.,][]{2004ApJ...600..149W,1999MNRAS.309..447M,2023ApJ...952..135C}. Observations in nearby galaxies indicate current tidal disruption rates of \(10^{-5}\sim 10^{-4}\,{\rm yr}^{-1}\) per galaxy, with a weak dependence on MBH mass for \(M_{\bullet} < 10^8\msun\)~\citep{2008A&A...489..543E,2018ApJ...852...72V,2020SSRv..216..124V}. 
The present-day tidal disruption rates in the explored M2G8 series models, as shown in Table~\ref{tab:model}, are consistent with these observational findings.}

{The derived present-day tidal disruption rates of very massive NSCs (\(M_{\rm cl}=10^9\msun\)), specifically models M2G91 and M2G91\_2, are \(\sim 10^{-3}\,{\rm yr}^{-1}\), which are relatively high. Possible explanations include: (1) Their current size remains much more compact than most NSCs in late-type massive galaxies, even after 12\,Gyr of evolution (see the left panel of Figure~\ref{fig:mbh_varying_ini}). Thus, these models may not represent the majority of massive galaxies observed today; (2) The mass growth of the MBH in these NSCs is significantly underestimated. For instance, the growth may not be dominated by loss cone accretion but by other mechanisms (e.g., stellar collisions, mass loss from stars, or gas streams accreted from the surrounding environment of NSCs). 
In reality, the MBHs in these massive NSCs may already have grown to \(>10^8\msun\) and, therefore, may no longer produce visible flares from tidal disruption events.
}

{For less massive clusters (\(M_{\rm cl} \sim 10^6\msun\), M2G7 model series), the present-day tidal disruption rates are relatively lower, on the order of \(10^{-6}\,{\rm yr}^{-1}\). However, since these rates are decreasing functions of time, many of these NSCs exhibited higher rates of \(>10^{-5} \sim 10^{-4}\,{\rm yr}^{-1}\) at \(t \lesssim 2-3\) Gyr. Consequently, if these clusters formed recently, as suggested by the estimated size expansion (see Section~\ref{subsec:size_expansion_of_NSCs}), their tidal disruption rates could still be consistent with observational estimates.
}

\subsubsection{Growth of the influence radius}

{There are two standard definitions of the influence radius of the MBH. The first is given by $r_h = \bh/\sigma^2$, where $\sigma$ is the one-dimensional stellar velocity. This quantity is generally not well-defined, as it can vary depending on the radius. However, it is common practice to assume an isothermal distribution of stars near the MBH, i.e., $\rho_\star \propto r^{-2}$, so that $\sigma$ is a constant value that can be equated to the velocity dispersion provided by the $\bh-\sigma$ relation—typically inferred from the bulge of the galaxy and measured far from the MBH. According to \citet{2013ARA&A..51..511K}, where $\sigma \propto \bh^{4.42}$, it is expected that $r_h \propto \bh^{0.55}$. This definition of $r_h$ has been widely used in many previous studies~\citep[e.g.,][]{Alexander05,Alexander09,2005ApJ...629..362H,2006ApJ...645L.133H,2013ApJ...764...52B,2004ApJ...600..149W,ZA24,2022hgwa.bookE..17A}.}

{Another standard definition is defined as the radius within which the enclosed mass 
equals to two times of the mass of the MBH, i.e., $M_\star(<r_h) = 2\bh$. 
The first definition of $r_h$ equal to the second one only if the density of stars around MBH is 
indeed isothermal. 
However, it is possible that they do not match with each other, e.g., in the case if the density distribution  deviates significantly from isothermal, or when 
the mass of MBH is much smaller than that of the NSCs~\citep{2011ApJ...735...89L,2017ARA&A..55...17A}.
As in our simulations the seed MBH is small, i.e., $10^3\sim 10^4\,\msun$, 
it is thus interesting for us to investigate whether the evolution 
$r_h$ from the enclosed mass aligns with the $r_h$ expected from the empirical $\bh-\sigma$ relation.}

The top left panel of Figure~\ref{fig:mbh_varying_ini_rh} shows the evolution of $r_h$ and $\bh$ in different models. Initially, the seed MBH has a mass of $10^3\sim 10^4\,\msun$, with an influence radius of about $0.01\sim 0.3$ pc. As the MBH mass increases and the NSC expands, $r_h$ continues to grow. 

For models with stellar masses larger than or comparable to those of Milky Way NSCs ($M_{\rm cl} \gtrsim 10^7\,\msun$, i.e., M2G8 or M2G9 series models), the growth of $r_h$ alongside the increase in $\bh$ follows approximately $r_h \propto \bh^{\beta}$, where $\beta = 0.5\sim 0.6$, 
 roughly follows the expectations 
of $r_h\propto \bh^{0.55}$. Note that 
the initial value of $r_h$ is (for model M2G84) or is not (for other models) around the expectations from the $\bh-\sigma$ relation. 
{This is mainly because the stellar density quickly relax to the distribution 
$\rho_\star\propto r^{-\gamma}$ with $\gamma=1.2\sim -1.4$ (See the bottom right panel of 
Figure~\ref{fig:mbh_varying_ini_rh}), such that $M(<r)\propto r^{3-\gamma}$ and $\beta=\frac{1}{3-\gamma}\simeq 0.5\sim 0.6$.}

Once the MBH mass exceeds approximately 10\% of the NSC mass, or when $r_h/r_{\rm eff}\gtrsim 0.1$ (See 
Top right panel of Figure~\ref{fig:mbh_varying_ini_rh}), the influence radius enclose 
the outer parts of the cluster where the density profile drops steeply, i.e., $\propto r^{-4}$.
As a result, $r_h$ increases faster as the mass of MBH growing ($\beta>0.6$). 
For lighter NSCs (e.g., M2G71-74, with masses around $10^6\,\msun$), the increase 
in $r_h$ is even more rapidly, as it is additionally driven by the rapid 
expansion of the size of the NSC.

Interestingly, from the top left panel of Figure~\ref{fig:mbh_varying_ini_rh}, we find that the majority of Milky Way-sized NSCs at 12 Gyr have $r_h$ values close to those expected from the empirical $\bh-\sigma$ relation. However, $r_h$ values for smaller NSCs are much larger than those defined by the $\bh-\sigma$ relation. These results suggest that the $r_h$ values defined by the  $\bh-\sigma$ relation, particularly when $\bh$ is small, may significantly underestimate the true influence radius in the cluster.

\subsubsection{Growth of the MBH's mass}
The final mass of MBHs strongly depends on the initial mass ($M_{\rm cl}(0)$) and size ($r_{\rm eff}(0)$) of the NSC (or the initial density in the inner regions). In Table~\ref{tab:model} we can see that, given the same mass of cluster, the smaller the initial size of the NSC, the greater it can grow to over cosmic time. Based on the initial $M_{\rm cl}(0)$ and $r_{\rm eff}(0)$ inferred from~\citet{2016MNRAS.457.2122G}, it appears challenging for NSCs to grow an MBH to a mass $\gtrsim 6\times 10^7\,\msun$, even for the most massive NSCs. This suggests that loss-cone accretion alone is insufficient to explain the formation of MBHs with masses $\gtrsim 6\times 10^7\,\msun$.

{For massive clusters, an initially higher slope index in the density profile at the inner regions can result in a slightly larger final MBH mass. For instance, M2G91\_2 has a final MBH mass of \(6\times10^6\msun\), which is slightly larger than its counterpart, M2G91, with a final mass of \(5\times10^6\msun\). However, for Milky Way-sized or smaller NSCs, increasing the initial slope index of the inner density profile leads to only a very small or negligible increase in the final MBH mass.}

For Milky Way-sized NSCs, it appears feasible to form MBHs with masses in the range of $10^5-10^7\,\msun$ over cosmic time purely through loss-cone accretion. However, for smaller NSCs with $M_{\rm cl}(0) \sim 10^6\,\msun$, mass growth is slower, and the final MBH mass typically falls within the range of $10^4-10^5\,\msun$. {The final mass of MBH 
 is only slightly smaller if we reduce the initial mass of the seed MBH from $10^4\,\msun$ to $10^3\,\msun$; i.e., 
 by comparing the result of model M2G82 with M2G82m (or M2G71 with M2G71m). This is because it usually takes only 
 about $\sim 0.2$ Gyr for a seed MBH grow from $10^3\,\msun$ to $10^4\,\msun$, which is only a small fraction of Hubble time.}

\subsubsection{Density distribution of stars and SBHs}

The right panel of Figure~\ref{fig:mbh_varying_ini_rh} shows the slope index of the density ($\gamma=d\ln \rho/d\ln r$) for stars and SBHs at present, averaged over the inner region of the cluster, specifically within $10^{-3}<r/r_{\rm eff}<0.1$. We find that the slope index $\gamma$ for stars in most models is relatively flat, ranging from $\gamma=-1.1$ to $-1.4$, which is consistent with observations of the Galactic center~\citep[e.g.,][]{2003ApJ...594..812G,2019MNRAS.484.1166M}. In contrast, the slope index for the density of SBHs is much steeper, with values of $\gamma=-2.2$ to $-1.75$. 

It is important to note that this result is only weakly dependent on the mass of the MBH, as it primarily reflects the effects of mass segregation, which influence the dynamics of the cluster regardless of the presence of the MBH.

\subsubsection{Comparing with observations of the Milky-Way NSC}

\begin{figure*}
	\center
	\includegraphics[scale=0.7]{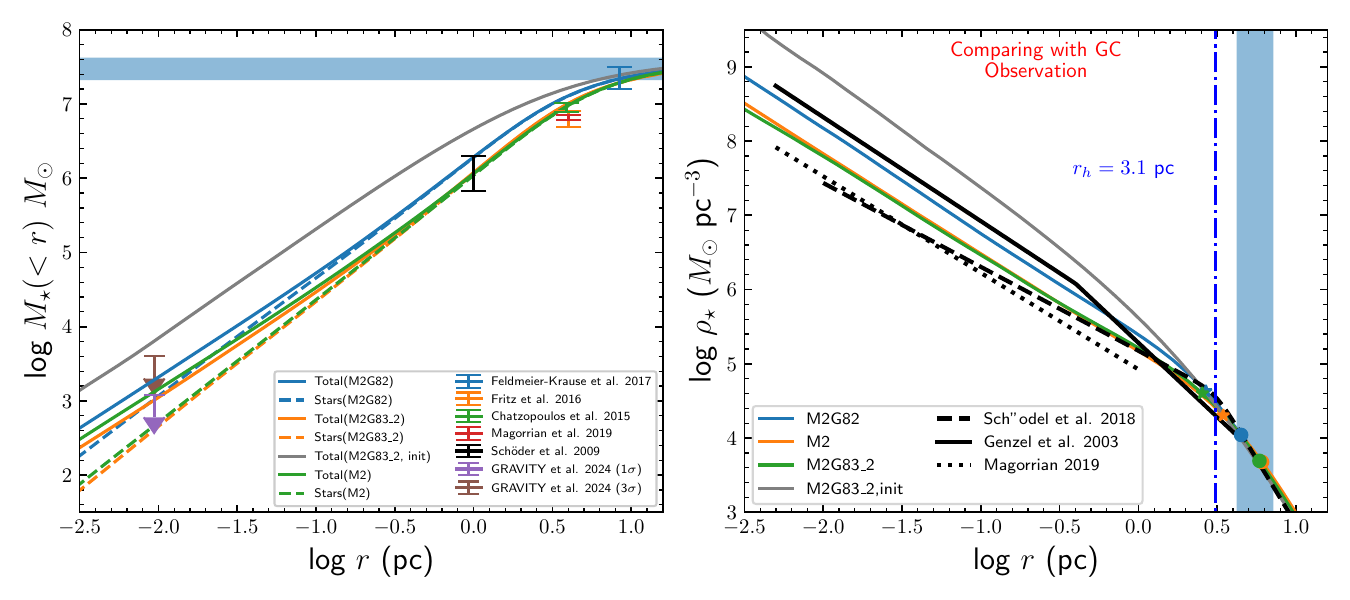}
	\caption{Left panel: Comparison of the enclosed mass in the model {M2G82\_2, M2G83 at $12$ Gyr (model M2 at $1T_{\rm rlx}$)} with those observed in Milky-Way NSC
	\citep{2017MNRAS.466.4040F,2016ApJ...821...44F,2015MNRAS.447..948C,2019MNRAS.484.1166M,2009A&A...502...91S}.
	{The upper limits show the $1\sigma$ ($\sim1200\msun$) and $3\sigma$ ($\sim 4000\msun$) constraints of the enclosed mass within 
	the orbit of S2~\citep{2024A&A...692A.242G}.}
	The light blue shaded region marks the constraints of the total stellar 
	mass,i.e., $2.1\sim 4.2\times 10^7\,\msun$
	\citep{2014A&A...566A..47S,2014A&A...570A...2F,2015MNRAS.447..948C,2016ApJ...821...44F,2017MNRAS.466.4040F};
	Right panel: Comparison of the stellar mass density profile at $12$ Gyr for model M2G82 and M2G83 
	with those observed 
	of the Milky-Way NSC by~\citet{2003ApJ...594..812G},~\citet{2018A&A...609A..27S}
	and~\citet{2019MNRAS.484.1166M}. The filled star and circle symbols are the positions of $r_h$ and $r_{\rm eff}$, respectively. 
	The shaded light-blue region marks the constraint of the effective radius of Milky-Way 
	NSC, i.e., $4.2\sim 7.2$pc~\citep{2014A&A...566A..47S,2016ApJ...821...44F}.
}
\label{fig:dehnen_gc}
\end{figure*}

It is interesting to compare the stellar density profile and the present-day MBH masses of the simulated models in \GNCc~with recent observations of our Milky Way's NSC. The current mass estimates for the Milky Way NSC are around $2\sim 4\times 10^7\,\msun$~\citep{2014A&A...566A..47S,2014A&A...570A...2F,2015MNRAS.447..948C,2016ApJ...821...44F,2017MNRAS.466.4040F}, so we focus primarily on models M2G8 series, which have an initial stellar mass of $4\times 10^7\,\msun$. Among these, models M2G82, {M2G82m, M2G83 and M2G83\_2} show the best consistency with observations. {As the results from M2G82m (M2G83) is quite similar to those of M2G82 (M2G83\_2), 
in the following we discuss mainly about model M2G82 and M2G83\_2. The comparisons between observations and those from M2G82 and M2G83\_2 are} illustrated in Figure~\ref{fig:dehnen_gc}.

In model M2G82, the present-day MBH mass is approximately $\bh \sim 4.1\times 10^6\,\msun$, which is consistent with the current estimate for the MBH mass in the Milky Way, i.e., $4.3\times 10^6\,\msun$~\citep{2009ApJ...692.1075G,2017ApJ...837...30G}. The stellar density profile in this model also aligns more closely with the stellar densities estimated by~\citet{2003ApJ...594..812G}. Model {M2G83\_2}, on the other hand, has a smaller MBH mass at 12 Gyr ($\bh \sim 2.8\times 10^6\,\msun$) and a stellar density profile more consistent with observations from~\citet{2018A&A...609A..27S} and~\citet{2019MNRAS.484.1166M}. In both models, the enclosed mass distribution, as shown in the left panel of Figure~\ref{fig:dehnen_gc}, is roughly consistent with observations of the Galactic center~\citep{2017MNRAS.466.4040F,2016ApJ...821...44F,2015MNRAS.447..948C,2019MNRAS.484.1166M,2009A&A...502...91S}. {The enclosed mass within the orbit of S2 is consistent with recent observations of the Galactic Center within \(3\sigma\) confidence levels~\citep{2024A&A...692A.242G}. It is important to note that this consistency arises from dynamical evolution, as the initial enclosed mass distribution can deviate significantly from the observed values (see the gray line in both panels of Figure~\ref{fig:dehnen_gc}, which represents the initial enclosed mass or density distribution for model M2G83\_2).
}

At 12 Gyr, the gravitational influence radius $r_h$ for M2G82 and {M2G83\_2} is 2.7 pc and {$2.5$} pc, respectively. 
the $r_h$ value in M2G82 is more consistent with the observed value of 
$3.1\sim 3.8$pc in our galactic center~\citep{2015MNRAS.447..948C,2018A&A...609A..27S}. 

The current effective radius is $r_{\rm eff} = 4.6$ pc ($r_{\rm eff} = 5.9$ pc) for M2G82 ({M2G83\_2}), which has grown from an initial value of $r_{\rm eff} \simeq 2.7$ pc ($r_{\rm eff} = 4.1$ pc). The current values for these models are consistent with those from recent observations, i.e., $r_{\rm eff} = 4.2\sim 7.2$ pc~\citep{2014A&A...566A..47S,2016ApJ...821...44F}. 

Additionally, both models have a present-day tidal disruption rate of stars of approximately $10^{-4}$ yr$^{-1}$, which aligns with theoretical expectations from steady-state solutions~\citep[e.g.,][]{1999MNRAS.309..447M,2004ApJ...600..149W,ZA24}.

{Model M2G82 adopts a seed MBH of $10^4\,\msun$. On the other hand, if we start from a seed black hole of $10^3\,\msun$, model M2G82m, the final mass of MBH is only slightly smaller, i.e., $3.8\times 10^6\,\msun$. 
We find that if we slightly reduce the initial size of the NSC to 
$\sim 2.5$ pc, the final mass of MBH can grow up to $4\times 10^6\,\msun$. 
}

{ Figure~\ref{fig:dehnen_gc} also shows that the enclosed mass distribution of model M2 at \(1T_{\rm rlx}\simeq 9\) Gyr is quite similar to that of model M2G83\_2. Models M2 and M2G83\_2 share similar initial and final effective radii for the cluster. This result suggests that, given sufficient evolutionary time, models with evolving MBHs may converge toward a quasi-steady-state solution with a similar effective radius. 
It is also noteworthy that model M2 demonstrates consistency with observational data. Within the orbit of S2, the enclosed mass in model M2 is \(\sim 1300\msun\), slightly exceeding the \(1\sigma\) upper limit suggested by \citet{2024A&A...692A.242G} (see also a similar model with five components simulated by \GNCc~in~\citet{2024A&A...692A.242G}).
}

In summary, for the Milky Way NSC, if it grew from a seed black hole of $10^3\sim 10^4\,\msun$ and initially had an effective radius about 1.3 (1.7) times smaller than its current value, the final mass of MBH can grow up to approximately 50\% (100\%) of the observed mass of MBH in the Galactic Center. In other words, it is possible to ascribe $\gtrsim$50\% of the current mass of MBH in Galactic center to the loss-cone accretion, 
if the NSC is initially $\gtrsim 30\%$ more compact than the current size. In this case, the simulated present-day effective radius and enclosed mass distribution are roughly consistent with the observations of the Galactic Center. Furthermore, the slope index of the density in the inner regions for these two models is $-1.2\sim -1.3$, which is also consistent with observations~\citep[e.g.,][]{2003ApJ...594..812G,2019MNRAS.484.1166M} (see the bottom right panel of Figure~\ref{fig:mbh_varying_ini_rh} or the right panel of Figure~\ref{fig:dehnen_gc}).

\subsubsection{The mass ratio of MBH to NSC}

The right panel of Figure~\ref{fig:mbh_varying_ini} shows the mass ratio between the MBH ($\bh$) and the NSC ($M_{\rm cl}$) after 12 Gyr of simulation. Generally, the ratio $\bh/M_{\rm cl}$ in most models falls within the range of $0.01-0.1$, which is consistent with observations from various studies~\citep{2020A&ARv..28....4N}. {These results are similar to those results in 
in~\citet{2004MNRAS.352..655A}, where they found that within a Hubble time 
a seed black hole of $\bh/M_{\rm cl}\sim 10^{-4}\sim 10^{-3}$ can eventually grow up to 
$\sim 10\%$ of the mass of the cluster.}

However, it is noteworthy that observations have revealed many NSCs with a central MBH that is much more massive than the total stellar mass of the NSC, with ratios of $\bh/M_{\rm cl} = 1\sim 10$. In principle, such high mass ratios could be achieved by simulating NSCs with a very small initial effective radius. However, when the MBH mass becomes significantly larger than the stellar mass of the NSC, the bulge components may start to influence the subsequent evolution. This influence may not be accurately captured in our simulations, as we assume that NSCs are in isolation. Therefore, we do not simulate or discuss these cases here but instead defer them to future studies.

\section{Conclusions and discussions}
\label{sec:conclusion_discussion}

In a previous work~\citep{ZA24}, we introduced a Monte Carlo method named \GNC~to study the dynamical evolution of a nuclear star cluster (NSC) composed of multiple mass components. However, in the earlier version, the effects of the stellar potential were neglected, limiting its applicability to NSCs. Here, we present a major update to \GNC~by incorporating the stellar potential and adiabatic invariant theory, enabling us to study the self-consistent dynamics of NSCs with a central massive black hole (MBH). We then use the updated version of \GNC~to investigate the cosmological evolution of NSCs and MBHs, assuming that the MBH grows its mass through the accretion of stellar objects falling within the loss cone, including the 
tidally disrupted stars and those of the non-flaring events, i.e., the direct swallowing of stars within the horizon 
or stellar-mass black holes falling within the last stable orbit. We also compare the present-day results of stellar densities, NSC sizes, and MBH masses from \GNC~with observations of nearby NSCs.

We conduct several tests on the new version of \GNC. When two-body dynamics are ignored, the updated code accurately reproduces the expected self-adjustment of the cluster under adiabatic invariant theory as the MBH mass gradually increases. We also tested the Plummer core collapse, finding that it results in a core collapse time of approximately $\sim 17t_{\rm rh}$, where $t_{\rm rh}$ is the half-mass relaxation time. Comparing this with the previous version of \GNC, which did not include the stellar potential, we observe good consistency in the inner parts of the cluster. However, the updated version shows significant differences in the outer parts, where the stellar potential dominates over that of the MBH. Importantly, the new version of \GNC~yields time-dependent dynamics of the NSC, rather than solutions that approach steady states.

{We apply \GNC~to study the cosmological evolution of NSCs and investigate how the mass of MBH evolves
 under the effect of loss-cone accretion alone (e.g., tidal disruption of stars and direct swallowing of stars within the event horizon or stellar-mass black holes within the last stable orbit). 
 As the first application of \GNC, we primarily aim to investigate the 
 contribution of the loss cone accretion to the total mass of MBHs in currently observed near-by galaxies, 
 although we notice that other complexities, such as stellar collisions and accretions of interstellar 
 gaseous materials, may also play important roles in mass growth of MBHs.}
 
{We ran such simulations over a Hubble time, i.e., 12 Gyr, starting with a seed MBH mass of $10^3\,\msun$ or $10^4\,\msun$. 
Our findings indicate that, for Milky Way-like NSCs (with stellar mass $M_{\rm cl}\sim 4\times 10^7\,\msun$), the MBH mass can grow to $10^5\sim 10^7\,\msun$ over 12 Gyr through pure loss-cone accretion, with the present-day effective sizes roughly consistent with current observations. For smaller NSCs with stellar mass $M_{\rm cl}\sim 10^6\,\msun$, the final MBH mass is usually $\lesssim 10^5\,\msun$, and these clusters experience significant size expansion over cosmic time. Based on current observations of NSC size and mass, we find that pure loss-cone accretion alone struggles to account for MBHs with masses greater than $6\times 10^7\,\msun$, even in the most massive NSCs ($M_{\rm cl}\sim 10^9\,\msun$). Thus, a significant portion of the MBH's mass may instead be contributed by other accretion channels, such as gas accretion~\citep{1982MNRAS.200..115S}.}

We observe that the influence radius $r_h$ of the MBH for Milky Way-sized NSCs grows in accordance with the $\bh-\sigma$ relation. However, for smaller NSCs, the relation is not well followed, primarily due to the rapid expansion of NSC sizes. Regardless of the central MBH mass, the current slope index of the stellar density is $-1.4\sim -1.1$, and for stellar-mass black holes, it is $-2.2\sim -1.75$, across various initial conditions.

{In the scenario of pure loss-cone accretion, }
the majority of the MBH's mass growth is due to the tidal disruption of stars. The cosmological evolution of the tidal disruption rate depends on the initial conditions of the NSC and the growth history of the MBH. Generally, the current tidal disruption rate is $\sim 10^{-4}$ yr$^{-1}$ for NSCs with mass $M_{\rm cl}\sim 10^7-10^8\,\msun$ and $10^{-7} \sim 10^{-6}$ yr$^{-1}$ for NSCs with mass $M_{\rm cl} \sim 10^6\,\msun$. The mass contribution from stellar-mass black holes falling into the loss cone is typically one or two orders of magnitude smaller than that from stars.

{Our results suggest that, starting from a seed black hole of $10^3\,\msun$ or $10^4\,\msun$, 
the mass of MBH after $12$ Gyr can grow up to $50\%$ ($100\%$) of the {Milky Way} MBH mass, 
if the NSC's effective radius was about 1.3 (1.7) times smaller than its current value in the early universe. 
In that case, and after $12$ Gyr of evolution, our simulations yield present-day mass distribution and effective radius consistent with
the observations of the Milky Way's NSC.
}

These findings have important implications for understanding how dynamical processes shape the size and structure of NSCs over cosmic time and the role of tidal disruption of stars in the growth history of central MBHs. We will present elsewhere the results of \GNC~incorporating additional factors crucial for cosmological evolution, in particular stellar collisions and, in the future, mergers, stellar evolution, star formation, and other gaseous materials that can be accreted by the MBH~\citep[e.g.,][]{1990ApJ...356..483Q,1991ApJ...370...60M,2002A&A...394..345F}, to develop a comprehensive study of the evolutionary history of NSCs and MBHs.

\section{Acknowledgments}
\noindent
We thank Eugene Vasiliev for helpful discussions and his kindness in providing the data for us.
We also thank Lu Youjun, Chen Yunfeng and Wang Long for helpful discussion of the results of the paper.
This work was supported in part by National Natural Science Foundation of 
China under grant No. 12273006,  11833007. 
This work was also supported in part by the Key Project of the National Natural Science Foundation 
of China under grant No. 12133004. The simulations in this work are performed partly in the TianHe II National
Supercomputer Center in Guangzhou. We acknowledge the funds from the ``European Union NextGenerationEU/PRTR'', Programa de Planes Complementarios I+D+I (ref. ASFAE/2022/014).


\end{document}